\documentclass[aps,prb, twocolumn,superscriptaddress,showpacs,longbibliography]{revtex4-1}

\usepackage[utf8]{inputenc}
\usepackage[english]{babel}
\usepackage{graphicx}
\usepackage{bm}
\usepackage{amsmath,amssymb,latexsym}
\usepackage{float}
\usepackage{xcolor}
\usepackage{tcolorbox}

\newcommand{\sign}{\text{sgn}}

\begin{document}

\title{On the dynamics of curved dislocation ensembles}
\author{István Groma}
\affiliation{Department of Materials Physics, E\"otv\"os Lor{\'a}nd University, P\'azm\'any P. stny. 1/A, 1117 Budapest, Hungary}
\author{Péter Dusán Ispánovity}
\affiliation{Department of Materials Physics, E\"otv\"os Lor{\'a}nd University, P\'azm\'any P. stny. 1/A, 1117 Budapest, Hungary}
\author{Thomas Hochrainer}
\affiliation{Institut für Festigkeitslehre, Graz University of Technology, Kopernikusgasse 24/I, 8010 Graz, Austria}
\begin{abstract}
To develop a dislocation-based statistical continuum theory of crystal plasticity is a major challenge of
materials science.
During the last two decades such a theory has been developed for the time evolution of a system of parallel edge dislocations. The evolution equations were derived by a systematic coarse-graining of the equations of motion of the individual dislocations and later retrieved from a functional of the dislocation densities and the stress potential by applying the standard formalism of phase field theories.
It is, however, a long standing issue if a similar procedure can be established for curved dislocation systems. An important prerequisite for such a theory has recently been established through a density-based kinematic theory of moving curves. In this paper, an approach is presented for a systematic derivation of the dynamics of systems of curved dislocations in a single slip situation. In order to reduce the complexity of the problem a “dipole” like approximation for the orientation dependent density variables is applied. This leads to a closed set of kinematic evolution equations of total dislocation density, the GND densities, and the so-called curvature density. The analogy of the resulting equations with the edge dislocation model allows one to generalize the phase field formalism and to obtain a closed set of dynamic evolution equations.

\end{abstract}

\maketitle

\section{Introduction}
Plastic deformation of crystalline materials is largely controlled by the motion of dislocations, that are line-type topological lattice defects.
Since the typical dislocation density in deformed metals is in the order of at least $\rho \sim 10^{14}$ m$^{-2}$, the average spacing between dislocation lines is less than 100 nm. This means that already micron sized samples contain a vast amount of strongly interacting dislocations. As a consequence, to model the plastic deformation of crystalline materials in terms of dislocations one has to handle the problem with statistical physics methods. However, there are two caveats for the direct application of methods from statistical physics to dislocation systems: (i) dislocation motion is strongly dissipative, and (ii) dislocations are flexible lines, inhibiting their treatment as point-particles.

The development of a statistical continuum theory of dislocations was initially motivated by the occurrence of size-effects \cite{fleck1997advances} in the plastic response of samples with characteristic dimensions on the order of 10 $\mu$m or less. Attempts to incorporate internal length scales into phenomenological continuum theories by considering so-called strain-gradients \cite{zhu1997strain,aifantis1999,fleck,gurtin}, did not yield a satisfying solution for general loading cases. Another key issue to be addressed is the ubiquitously observed dislocation pattern formation during plastic deformation. Since the early 1960s several theoretical and numerical attempts have been suggested, initially based on analogies with other physical problems like spinodal decomposition \cite{holt1970}, internal energy minimization \cite{hansen1986}, or chemical reaction-diffusion systems \cite{aifantis1985,pontes}. Since, however, they are not directly linked to the specific properties of individual dislocations they are fundamentally phenomenological approaches. Dislocation patterning was also an important motivation for the development of the discrete dislocation dynamics (DDD) method \cite{kubin1992,ghoniem1999,devincre2001,devincre2002}. But due to the long range dislocation-dislocation interaction the simulations are computationally extremely expensive and the study of dislocation patterning with DDD is still limited to specific problems like irregular clusters or veins \cite{devincre2001,devincre2002,hussein2015}. Recently, El-Azab and coworkers \cite{xia2015computational,lin2020implementation} used a continuum formulation based on vector dislocation densities in large-scale numerical simulations, which seem to feature the evolution of dislocation patterns. However, this pseudo-continuum variant of DDD is a numerical rather than a theoretical model of dislocation patterning. 

The two caveats for developing a statistical continuum theory of dislocations named in the first paragraph have been approached largely independently from each other so far. The consequences of the dissipative nature of dislocation motion has been thoroughly explored in strongly simplified quasi two-dimensional systems of straight parallel edge dislocations, where dislocations are treated as signed point particles moving in a plane. By a systematic coarse-graining of the evolution equations of individual dislocations  \cite{groma1997link,zaiser2001statistical,groma2003spatial,groma2007dynamics,mesarovic2010,dogge2015,groma2016dislocation,valdenaire2016density,groma2019statistical} a continuum theory was developed during the last 20 years, that has been successfully compared to discrete dislocation dynamics (DDD) simulations \cite{groma2003spatial,yefimov2004comparison,groma2006debye,ispanovity2020emergence}. By now, it can be considered as a well-established theory for the 2D problem it addresses. It was moreover shown that the model can be also formulated as a specific phase field theory \cite{groma2007dynamics,groma2010variational,groma2015scale,groma2016dislocation}. In contrast to many other phase field theories, the phase field functional in this case could be strictly derived from the statistical theory, and is not obtained on phenomenological grounds. The most important feature of the theory is that it predicts dislocation patterning although it was not ``designed'' for it \cite{groma2016dislocation,wu2018instability,ispanovity2020emergence}.

The fact that dislocations are moving flexible lines entails the question, what are suitable continuum variables allowing for a closed system of conservation laws for dislocation systems. This has been answered in a primarily kinematic theory of curved dislocations, which was developed by Hochrainer \emph{et al.}\cite{hochrainer2006fundamentals,hochrainer2007three,sandfeld2010numerical,hochrainer2014continuum,hochrainer2015multipole}. The kinematics were initially derived in a higher dimensional space, containing the line direction as independent variable. A multipole expansion of the theory leads to a formulation in terms of alignment tensors, which, in the case of only planar dislocations on parallel glide planes, is equivalent to a Fourier expansion. The resulting conservation laws may be used to derive ``kinetic'' theories from a thermodynamic potential with standard methods from irreversible thermodynamics. This yields forms of driving forces \cite{hochrainer2016thermodynamically}, naturally generalizing those found in the quasi-two-dimensional theory. A thermodynamic potential in terms of alignment tensors has been suggested by Zaiser \cite{zaiser2015local} based on a local density approximation of the interaction energy. However this potential has not been derived from the microscopic kinetics, and, though the form is very similar, when specialized to the straight dislocation case, it does not reproduce the potential derived in Ref.~\onlinecite{groma2016dislocation}.

In the current paper we provide a synthesis of the quasi-2D and the curved dislocation theory, by deriving the thermodynamic potential within a dipole-type Fourier approximation of the higher dimensional variables. In the first part of the paper the 2D continuum theory and the 2+1D theory of the kinematics of curved dislocations are shortly summarized. In the main part of the paper it is shown that within a dipole type approximation a closed thermodynamically consistent continuum theory of the evolution of curved dislocations can be established.

\section{2D dislocation dynamics}
\label{sec:2d}

Before we start to discuss the problem of the dynamics of curved dislocations let us shortly summarize the continuum theory of straight parallel edge dislocations. The main physical ideas presented here will serve as a basis for deriving generalized dynamic equations in the 3D case.

\begin{figure}[!ht]
\includegraphics[angle=0,width=8.5cm]{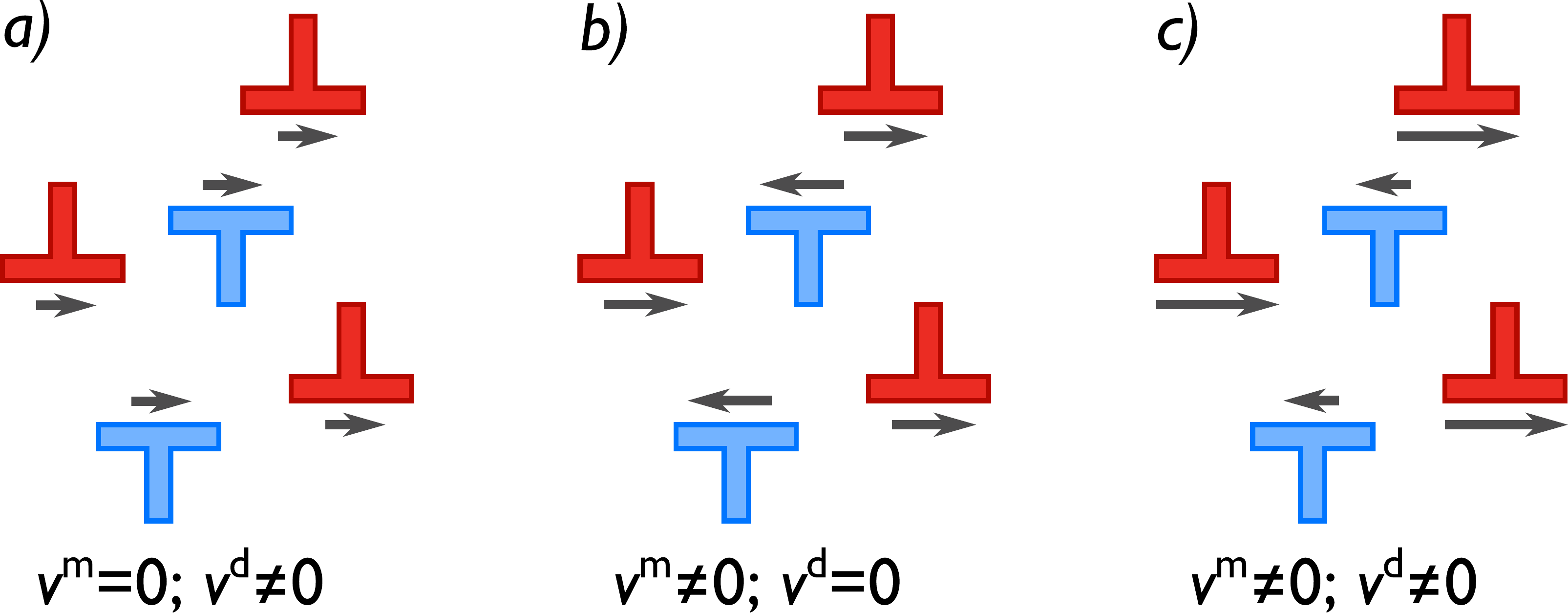}
\caption{(a): Sketch of situation when positive (red) and negative (blue) sign dislocations move locally with the same velocity and the relative velocity of the two types of dislocations is zero ($v^\mathrm m = 0$), so, the configuration moves as a ``rigid body''. This means that if $\kappa \ne 0$ then there is a corresponding non-zero plastic strain rate according to Eq.~\eqref{eq:gg}. (b): When $v_+ = -v_-$ then there are relative displacements between positive and negative dislocations but $v^\mathrm d = 0$. (c): A general case when both $v^\mathrm m$ and $v^\mathrm d$ are non-zero. The shown example is the superposition of panels a) and b).}
\label{fig:velocities}
\end{figure}

In this section we assume that dislocations are parallel with the $y$ axis of a Cartesian coordinate system and their Burgers vector points in the $x$ direction. In such a case we can distinguish two types of dislocations, ``positive'' ones with Burgers vector $(b, 0, 0)$, and ``negative'' ones with $(-b, 0, 0)$. Since dislocation positions can be characterized by their intersection point with the $xz$ plane, the problem is essentially 2D. The evolution of the system on the level of the densities of dislocations with different sign ($\rho_+$ and $\rho_-$) is described by balance equations that ensure conservation of the total number of dislocations of both type:
\begin{equation}
  \partial_t\rho_{\pm}+\partial_x[\rho_{\pm} v_{\pm}]= f(\rho_+,\rho_-),
\end{equation}
where $v_+$ and $v_-$ are the average velocities of the positive and negative dislocations in the slip plane, and $f(\rho_+,\rho_-)$ is a source term \cite{groma2016dislocation}. Since dislocations cannot be created or annihilated in the steady state of the system, the source term has to be proportional to the plastic deformation rate $\dot \gamma$, which is given by Orowan's law as
\begin{equation}
\dot\gamma=b(\rho_+ v_+-\rho_-v_-),
\end{equation}
so,
\begin{equation}
 f(\rho_+,\rho_-)=\dot\gamma \Psi(\rho_+,\rho_-)
\end{equation}
with an appropriate $\Psi$ function. By adding and subtracting the two equations one obtains
\begin{equation}
 \partial_t\rho+\partial_x[\rho v^\mathrm d+\kappa v^\mathrm m]=|\dot \gamma| \Psi(\rho,\kappa),
 \label{eq:rr}
\end{equation}
\begin{equation}
 \partial_t\kappa+\partial_x[\rho v^\mathrm m+\kappa v^\mathrm d]=0, \label{eq:kk}
\end{equation}
\begin{equation}
\dot\gamma=b(\rho v^\mathrm m+\kappa v^\mathrm d),
 \label{eq:gg}
\end{equation}
where $\rho=\rho_++\rho_-$ is the statistically stored dislocation (SSD) density, $\kappa=\rho_+-\rho_-$ is the geometrically necessary dislocation (GND) density, and 
$v^\mathrm m=(v_+-v_-)/2$ and $v^\mathrm d=(v_++v_-)/2$ are the ``mean'' and ``difference'' or ``drift'' velocities, respectively \cite{wu2018instability}. Figure \ref{fig:velocities} provides a sketch on the physical meaning of these quantities.

\begin{figure}[!ht]
\includegraphics[angle=0,width=5cm]{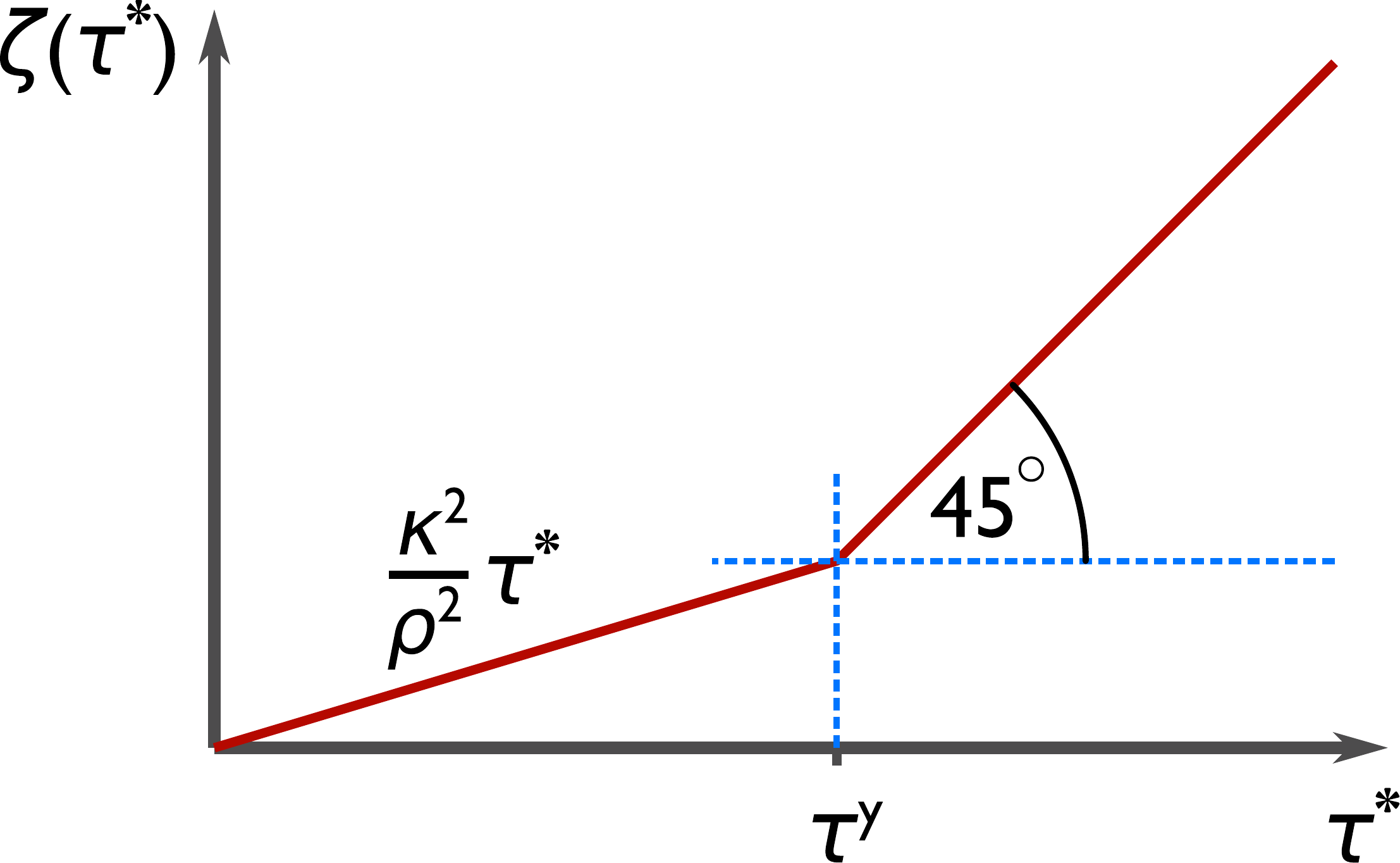}
\caption{The $\zeta(\tau^*)$ mobility function \cite{groma2016dislocation}}
\label{fig:mob0}
\end{figure}

Equations (\ref{eq:rr},\ref{eq:kk}) together with (\ref{eq:gg}) represent the kinematics of straight parallel edge dislocations. When constructing dynamic equations the question is how velocities $v^\mathrm m$ and $v^\mathrm d$ depend on the microstructure represented by the densities $\rho$ and $\kappa$. Previously Groma \emph{et al.}\ performed the systematic coarse graining of the equation of motion of individual dislocations to derive dynamic evolution equations for $\rho$ and $\kappa$ that read as\cite{groma2016dislocation}
\begin{equation}
 \partial_t \rho + M_0 b\partial_x\left(\kappa\tau^*+\rho \tau^\mathrm d\right) = 0,
 \label{eq:ev1}
\end{equation}
\begin{equation}
  \partial_t \kappa + M_0 b\partial_x\left(\kappa \tau^\mathrm d+\rho \zeta(\tau^*)\right) = 0, \label{eq:ev2}
\end{equation}
with the mobility function (see Fig.~\ref{fig:mob0}):
\begin{equation}
 \zeta(\tau^*)=\left\{ 
\begin{array}{ll}
      \frac{\kappa^2}{\rho^2}\tau^*, & {\rm if} \ |\tau^*| \leqq \tau^\mathrm y, \\
      \tau^* - s\tau^\mathrm y \left(1-\frac{\kappa^2}{\rho^2} \right), & {\rm if} \ |\tau^*| > \tau^\mathrm y, 
    \end{array}
 \right.
\label{eq:Mpm}
\end{equation}
where $s=\sign(\tau^*)$.
Here stress terms were introduced that can be calculated from $\rho$ and $\kappa$ and their spatial derivatives \cite{groma2016dislocation}. The term $\tau^*$ is the sum of the  ``mean-field'' stress $\tau^\mathrm{mf}$ and the ``back-stress'' $\tau^\mathrm b$:
\begin{equation}
    \tau^* = \tau^\mathrm{mf} + \tau^\mathrm b.
\end{equation}
The mean-field stress is the resolved shear stress in the glide plane due to the long-range stresses of the GNDs and the surface tractions and displacements. The back-stress $\tau^\mathrm b$ and the ``diffusion stress'' $\tau^\mathrm d$ read as
\begin{eqnarray}
    \tau^\mathrm b &=& -Gb \frac{D}{\rho} \partial_x \kappa, \\
    \tau^\mathrm d &=& -Gb A \, \partial_x \rho,
\end{eqnarray}
where $G=\frac{\mu}{2\pi(1-\nu)}$ is an elastic constant ($\mu$ and $\nu$ are the shear modulus and Poisson's ratio, respectively), $D$ and $A$ are dimensionless constants, and $\tau^\mathrm y=\alpha\mu b\sqrt{\rho}$ is the local yield stress with $\alpha$ being the dimensionless Taylor coefficient in accordance with the Taylor hardening law.

\begin{figure}[!ht]
\includegraphics[angle=0,width=5.5cm]{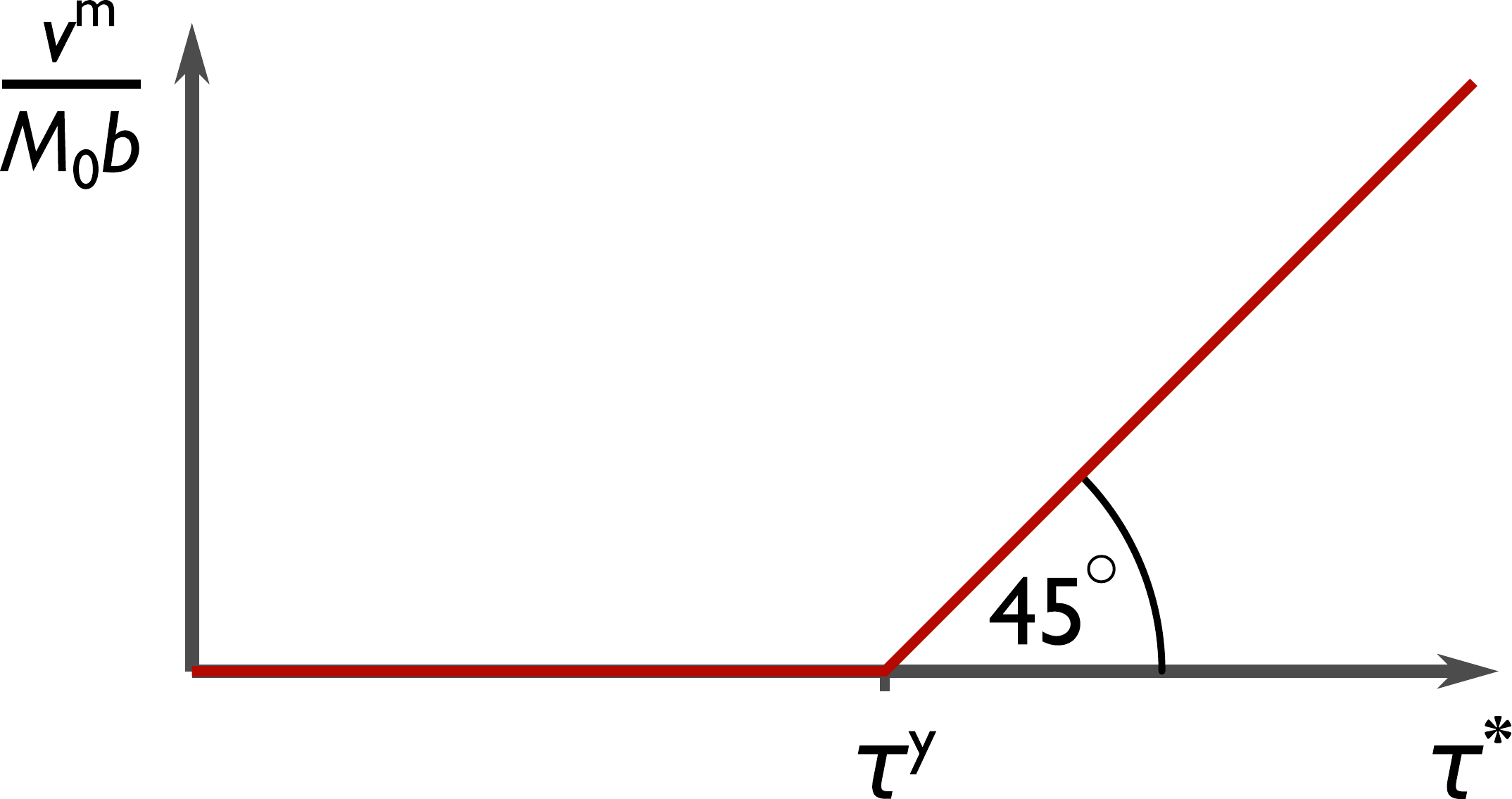}\\
\vspace*{0.5cm}
\includegraphics[angle=0,width=5.5cm]{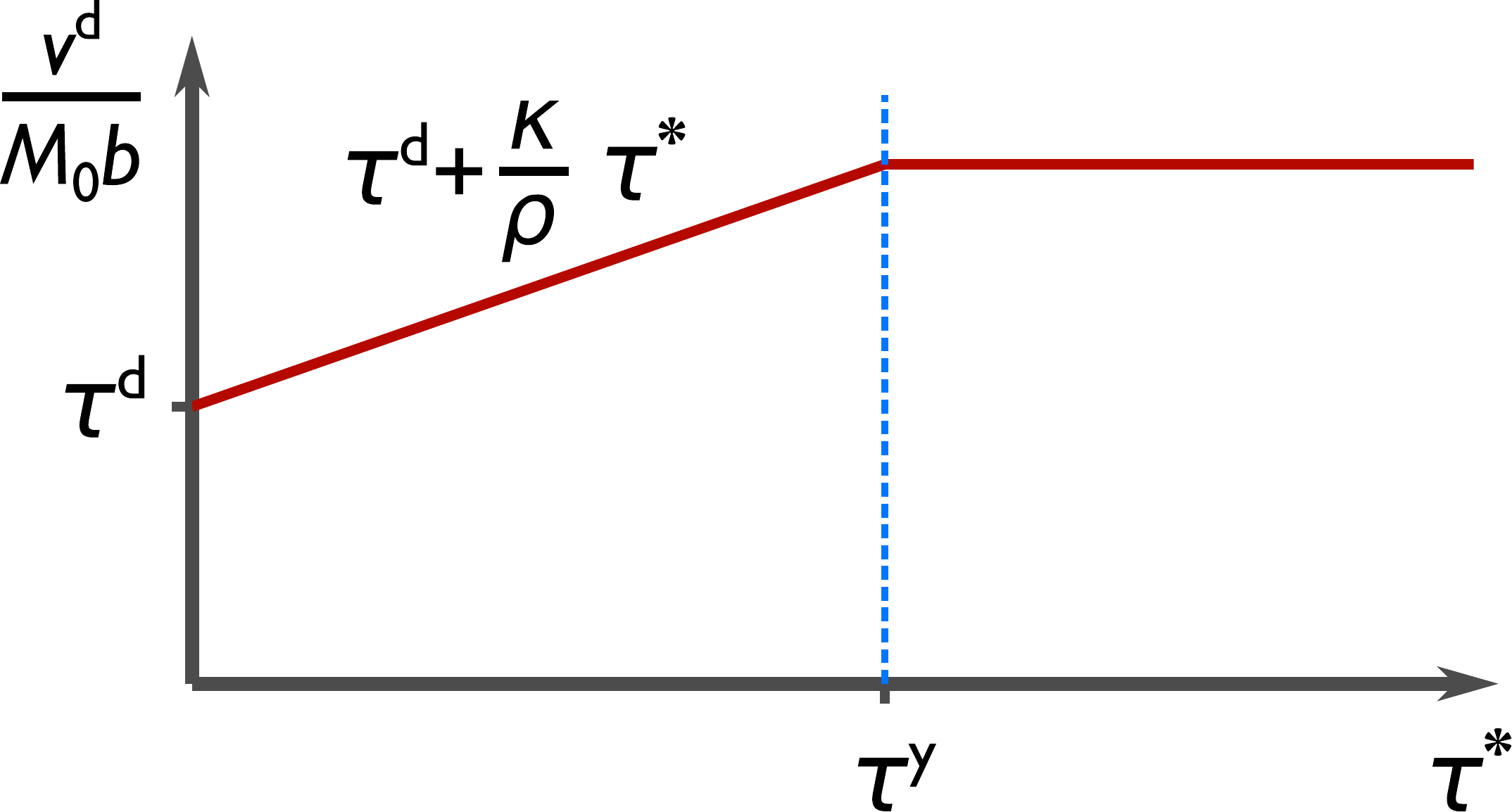}
\caption{The velocities $v^\mathrm m$ and $v^\mathrm d$ as a function of $\tau^*$ for the $\kappa >0$ case.} \label{fig:vel}
\end{figure}

In order to formulate these equations in terms of the mean and drift velocities ($v^\mathrm m$ and $v^\mathrm d$, respectively) introduced above from Eqs.~(\ref{eq:rr},\ref{eq:kk},\ref{eq:ev1},\ref{eq:ev2}) one obtains
\begin{eqnarray}
 \left(v^\mathrm d+\frac{\kappa}{\rho} v^\mathrm m\right)&=&M_0 b\left(\frac{\kappa}{\rho}\tau^*+\tau^\mathrm d\right), \label{eq:plus}\\
 \left(v^\mathrm m+\frac{\kappa}{\rho} v^\mathrm d\right)&=&M_0 b\left(\frac{\kappa}{\rho}\tau^\mathrm d+\zeta(\tau^*)\right). \label{eq:minus}
\end{eqnarray}
After rearranging these equations $v^\mathrm m$ and $v^\mathrm d$ can be expressed as
\begin{eqnarray}
 v^\mathrm m&=&M_0 b\chi (\tau^*), \label{eq:v0} \\
 v^\mathrm d&=&M_0b\left( \frac{\kappa}{\rho} (\tau^*-\chi(\tau^*)) +\tau^\mathrm d\right), \label{eq:v1}
\end{eqnarray}
where another mobility function $\chi$ was introduced that reads as
\begin{equation}
 \chi(\tau^*)=\left\{ 
\begin{array}{ll}
      0,& {\rm if} \ |\tau^*|\leqq\tau^\mathrm y, \\
      \tau^*-s\tau^\mathrm y,  & {\rm if} \ 
|\tau^*| > \tau^\mathrm y. 
    \end{array}
 \right.
\label{eq:chi}
\end{equation}
Equations (\ref{eq:v0},\ref{eq:v1}) together with Eqs.~(\ref{eq:rr},\ref{eq:kk}) form a closed set of evolution equations that are equivalent to Eqs.~(\ref{eq:ev1},\ref{eq:ev2}). Introduction of the $v^\mathrm m$ and $v^\mathrm d$, however, does not only yield equations that are mathematically somewhat simpler, but also highlight the physics behind the mobility laws. According to Fig.~\ref{fig:vel}, the mean velocity $v^\mathrm m$ is exactly zero up to the yield stress $\tau^\mathrm y$. This means that the relative positions of positive and negative dislocations with respect to each other do not change below the threshold stress $\tau^\mathrm y$ and dislocation configuration drifts as a ``rigid body''. Indeed, in this regime (that is, $|\tau^*|\leqq\tau^\mathrm y$) $v^\mathrm d$ may be positive if either $\tau^\mathrm d$ or $\kappa$ is non-zero. This situation is visualized in Fig.~\ref{fig:velocities}(a). Above the yield stress ($|\tau^*|>\tau^\mathrm y$) $v^\mathrm m$ becomes non-zero, that is, the configuration is no more ``rigid'', but rearrangements within the structure of positive and negative dislocations start to take place. The drift velocity $v^\mathrm d$ remains constant as seen in Fig.~\ref{fig:vel}. For a sketch of the corresponding dislocation velocities see Fig.~\ref{fig:velocities}(c).

It is important to point out that although dislocation system moves as a rigid body below $\tau^\mathrm y$ and no relative rearrangements take place, this does not necessarily imply a zero plastic strain rate. Indeed, according to Eq.~(\ref{eq:gg}), if in some region of the sample $\kappa \ne 0$, then there the drift velocity will lead to a non-zero strain rate even below the yield point. It may seem odd that dislocation motion and even a non-zero plastic strain rate can be observed below the yield point. We point out that $\tau^\mathrm y$ is a \emph{local} yield point and should not be confused with an emerging global yield stress. If the external stress (being part of $\tau^{\mathrm{mf}}$ and, consequently, of $\tau^*$) is below the global yield point then gradients in $\rho$ and $\kappa$ will develop that will stop dislocation motion and, thus, plastic strain rate will become zero everywhere in the material \cite{groma2016dislocation, wu2018instability}. 

The evolution equations presented so far were obtained by a systematic coarse graining of the discrete microstructure. The same equations, however, can be also derived using general thermodynamics principles. In the following we will review this procedure and its main conceptual steps, since this methodology can be more easily generalized to the 3D case. Note, that the term thermodynamic refers to the general conceptual framework being adopted from irreversible thermodynamics, but does not by any means refer to the role of temperature. In this paper we deal with systems where thermal noise does not play any role.

We start by noticing that due to the dissipative nature of the motion of dislocations (force action on a dislocation is proportional to the dislocation velocity) the total elastic energy of the dislocation system cannot increase during the evolution of the system. Accordingly, there exists a scalar quantity (state variable) for the discrete dislocation system that cannot increase as the system evolves (see Appendix D). 

As it was shown earlier in detail, for the coarse grained system there also exists a scalar functional of the dislocation densities $P[\rho,\kappa]$ that cannot increase during the evolution of the system\cite{groma2007dynamics,groma2010variational,groma2016dislocation}. This quantity was found no to be equal to the coarse grained elastic energy, so, the term ``plastic potential'' was introduced for it\cite{groma2006debye,groma2007dynamics,groma2010variational} (details are given in Appendix D). 

Let us see what are the consequence that $\dot{P}\leqq0$:
\begin{eqnarray}
 \dot P=\int \left[\mu_\rho \partial_t \rho+\mu_\kappa \partial_t \kappa \right] \mathrm dA,
\end{eqnarray}
where
\begin{eqnarray}
 \mu_\rho&=&\frac{\delta P}{\delta \rho}, \\
 \mu_\kappa&=&\frac{\delta P}{\delta \kappa}.
\end{eqnarray}

By substituting the general kinematic Eqs.~(\ref{eq:rr}) and (\ref{eq:kk}) into the above equation, after partial integration one gets 
\begin{equation}
\begin{split}
 \dot P=\int & \left\{
 (\rho v^\mathrm d+\kappa v^\mathrm m)\partial_x \mu_\rho + \mu_\rho|\dot \gamma| \Psi(\rho,\kappa) \right. \\ &+\left. (\rho v^\mathrm m+\kappa v^\mathrm d) \partial_x\mu_\kappa
 \right\} \mathrm dA.
\end{split}
\end{equation}
If we ensure that
\begin{eqnarray}
 \mu_\rho |\dot \gamma| \Psi(\rho,\kappa)<0 \label{eq:source}
\end{eqnarray}
we have to fulfill the condition
\begin{eqnarray}
 \tau^\mathrm d\left(v^\mathrm d+\frac{\kappa}{\rho} v^\mathrm m\right)+\tau^*
 \left(v^\mathrm m+\frac{\kappa}{\rho} v^\mathrm d\right) \geqq 0, \label{eq:positive}
\end{eqnarray}
where the notations
\begin{eqnarray}
 \tau^*&=&-\frac 1b \partial_x \mu_\kappa, \label{eq:tau_star} \\
 \tau^\mathrm d&=&-\frac 1b \partial_x \mu_\rho \label{eq:tau_d}
\end{eqnarray}
were introduced.

The next task is to find a mobility rule for $v^\mathrm m$ and $v^\mathrm d$ that fulfils condition (\ref{eq:positive}). There is no unique solution for this step so one needs to employ a physical argument. Here we refer to the derivation described earlier in this section and adopt mobility rules of Eqs.~(\ref{eq:v0},\ref{eq:v1}). This choice also explains the notations used in Eqs.~(\ref{eq:tau_star},\ref{eq:tau_d}): $\tau^*$ and $\tau^\mathrm d$ are equivalent quantities with the ones obtained during the coarse graining procedure. In the following it is shown that this choice for the mobility laws is indeed compatible with Eq.~(\ref{eq:positive}).

Since below the flow stress ($|\tau^*| \leqq \tau^\mathrm y$) 
\begin{eqnarray}
 v^\mathrm m&=&0 \\
 v^\mathrm d&=&M_0b\left( \frac{\kappa}{\rho}\tau^* +\tau^\mathrm d\right)
\end{eqnarray}
the inequality (\ref{eq:positive}) is trivially  fulfilled. 

In the flowing regime ($|\tau^*|>\tau^\mathrm y$) 
\begin{eqnarray}
 v^\mathrm m&=&M_0 b(\tau^*-s\tau^\mathrm y),\\
 v^\mathrm d&=&M_0b\left( s\frac{\kappa}{\rho}\tau^\mathrm y +\tau^\mathrm d\right).
\end{eqnarray}
So the inequality (\ref{eq:positive}) reads as
\begin{equation}
\begin{split}
& \left(\tau^*+\frac{\kappa}{\rho}\tau^\mathrm d\right)(\tau^*-s\tau^\mathrm y) \\
& \quad +  \left(\frac{\kappa}{\rho}\tau^*+\tau^\mathrm d\right)\left(s\frac{\kappa}{\rho}\tau^\mathrm y+\tau^\mathrm d\right) \geqq 0.
\end{split} 
\label{eq:ine2}
\end{equation}
By introducing 
\begin{eqnarray}
 B^\mathrm d=s\frac{\kappa}{\rho}\tau^\mathrm y+\tau^\mathrm d
\end{eqnarray}
Eq.~(\ref{eq:ine2}) can be reformulated as
\begin{equation}
\begin{split}
 &\left[\tau^*-s\tau^\mathrm y+\left(1-\frac{\kappa^2}{\rho^2}\right)s\tau^\mathrm y +\frac{\kappa}{\rho}B^\mathrm d\right](\tau^*-s\tau^\mathrm y) \\
 & \quad +\left(\frac{\kappa}{\rho}(\tau^*-s\tau^\mathrm y)+B^\mathrm d\right)B^\mathrm d \geqq 0
\end{split}
\label{eq:ine3}
\end{equation}
that can be rewritten as
\begin{equation}
\begin{split}
 & \begin{pmatrix}
       \tau^*-s\tau^\mathrm y, & B^\mathrm d \end{pmatrix}
 \begin{pmatrix}
        1  & \frac{\kappa}{\rho} \\
        \frac{\kappa}{\rho} & 1
\end{pmatrix}
\begin{pmatrix}
       \tau^*-s\tau^\mathrm y\\ B^\mathrm d
\end{pmatrix}
\\
& \quad +\left(1-\frac{\kappa^2}{\rho^2}\right)s\tau^\mathrm y(\tau^*-s\tau^\mathrm y)\geqq 0.
\end{split}
\end{equation}
Since the matrix in the first row is positive definite and $s\tau^\mathrm y(\tau^*-s\tau^\mathrm y) > 0$, the above form clearly indicates that the inequality (\ref{eq:positive}) is indeed fulfilled in the flowing regime too.  So, the mobility laws of Eqs.~(\ref{eq:v0},\ref{eq:v1}) do guarantee that the plastic potential $P$ cannot increase during the evolution of the system irrespective of the actual form of $P$. It should be noted that the condition (\ref{eq:source}) imposes an important restriction on the physically acceptable form of the source term introduced mostly on a phenomenological ground.

At the end of this section the main steps of the thermodynamic considerations are summarized. To arrive at a closed set of evolution equations, firstly, one needs to define the plastic potential $P$ as a function of the SSD and GND densities ($\rho$ and $\kappa$, respectively). Secondly, stress terms $\tau^*$ and $\tau^\mathrm d$ follow according to Eqs.~(\ref{eq:tau_star},\ref{eq:tau_d}). Thirdly, mean and drift velocities are obtained using the mobility rules of Eqs.~(\ref{eq:v0},\ref{eq:v1}). Finally, the evolution equations follow after substituting these into Eqs.~(\ref{eq:rr},\ref{eq:kk}). The actual form of the stress terms $\tau^*$ and $\tau^\mathrm d$ and that of the plastic potential $P$ are given in Appendix D, details can be found in  Refs.~\onlinecite{groma2016dislocation, groma2019statistical}.

In the following the generalization of the above results to curved dislocations will be discussed after recapitulating the kinematic fundamentals of the continuum theory of curved dislocations.

\section{Kinematics of curved dislocations}

For describing the kinematics of the evolution of curved dislocation ensembles we follow the method developed by Hochrainer \emph{et al.}\cite{hochrainer2006fundamentals,hochrainer2007three,sandfeld2010numerical,hochrainer2014continuum,hochrainer2015multipole}. For simplicity we consider only glide type dislocation motion with single slip (with slip plane perpendicular to the $z$ axis). We also assume that there are no dislocations in the other slip systems, i.e. we exclude forest dislocations. 

To describe the evolution of the dislocations we extend the problem into 2+1 dimensions \cite{hochrainer2006fundamentals,hochrainer2007three,hochrainer2014continuum}. The third dimension is the line direction, represented by the angle $\varphi$ the dislocation line direction forms with the $x$ axis, given by, e.g.,\ the Burgers vector. So, the problem is ``expanded'' to the $(x,y,\varphi)=(\bm{r},\varphi)$ space. The static state of the system is given by a vector field ${\bm R}$ on the 2+1D space characterized by two scalar fields \cite{hochrainer2006fundamentals,hochrainer2007three,hochrainer2014continuum}, the dislocation density $\rho'(\bm{r},\varphi)$, and the so-called curvature density $q'(\bm{r},\varphi)$ as
\begin{equation}
{\bm R}=(\rho' \cos \varphi, \rho' \sin \varphi, q')= \rho' {\bm L}
\end{equation}
with the line direction ${\bm L}$ given by
\begin{equation}
 {\bm L}=(\cos \varphi, \sin \varphi, k')=(\bm{l},k'),
\end{equation}
where $k' = q'/ \rho' $ is interpreted as the local average curvature \cite{hochrainer2006fundamentals,hochrainer2007three} of the dislocations with spatial line direction $\bm{l}$. In this paper scalar functions having domain in the 2+1D space are distinguished by the $(\cdot)'$ sign. Equivalent to the line direction ${\bm L}$ we introduce the operator
\begin{equation}
 \hat{L}=\cos \varphi\, \partial_x+\sin \varphi \, \partial_y+ k \partial_\varphi. \label{eq:L}
\end{equation}
It is important to mention that the fields introduced certainly can depend on the $z$ coordinate too, but since for the geometry considered the loop evolution happens in a plane perpendicular to the $z$ axis, the $z$ dependence is not indicated until it is not necessary.

A generalization of the dislocation density tensor is given through the ``signed line density'' on the 2+1D space as
\begin{equation}
 {\bm \alpha}'=\rho'(\bm{r},\varphi){\bm L}(\bm{r},\varphi)\otimes \bm{b}={\bm R}\otimes \bm{b}
\end{equation}
where $\bm{b} = (b_x,b_y)$ is the Burgers vector of the dislocation loops considered. 
The dislocation density tensor in the 2D ``real'' space is the average of the spatial part of ${\bm \alpha}'$ in the $\varphi$ direction,
\begin{equation}
 \bm \alpha(\bm{r})=\frac{1}{2\pi}\int_0^{2\pi}\rho'(\bm{r},\varphi){\bm{l}}(\bm{r},\varphi)\otimes \bm{b} \; \mathrm d\varphi =\bm{\kappa} \otimes \bm{b}.
 \label{eq: def alpha}
\end{equation}
Because the Burgers vector $\bm{b}$ is independent of the line direction, the dislocation density has a product structure with the net line-direction vector $\bm \kappa = (\kappa_1, \kappa_2)$, the components of which are the first order Fourier coefficients of $\rho'$,
\begin{eqnarray}
     \kappa_1(\bm r) &=& \frac 1{2\pi} \int_0^{2\pi} \rho'(\bm{r},\varphi) \cos{\varphi} \, \mathrm d \varphi, \label{eq:kappa1} \\
     \kappa_2(\bm r) &=& \frac 1{2\pi} \int_0^{2\pi} \rho'(\bm{r},\varphi) \sin{\varphi} \, \mathrm d \varphi. \label{eq:kappa2}
\end{eqnarray}
If the angle $\varphi$ is taken from the Burgers vector, $\kappa_1$ is the net screw dislocation component and $\kappa_2$ the net edge dislocation component.

It is important to ensure that dislocation lines do not end in the system. We force this condition in the 2+1D space (i.e. we do not allow discontinuity in the $\varphi$ direction) \cite{hochrainer2006fundamentals,hochrainer2007three,hochrainer2014continuum}. This is ensured by the condition:
\begin{equation}
 \hat{\text{Div}} \, {\bm \alpha}'=0 \label{eq:alpha}
\end{equation}
where the generalized operator $\hat{\text{Div}}$ acts on a vector field ${\bm A}=(A_1,A_2,A_\varphi)$ as
\begin{equation}
 \hat{\text{Div}}{\bm A}=\partial_x A_1+\partial_y A_2 + \partial_\varphi A_\varphi,
\end{equation}
leading to the condition:
\begin{equation}
 \cos \varphi \, \partial_x \rho'+\sin \varphi \, \partial_y \rho' + \partial_\varphi q'=0. \label{eq:noend}
\end{equation}
Note that the solenoidality of ${\bm \alpha}'$ implies the solenoidality of $\bm \alpha$ via Eq.~\eqref{eq: def alpha}.

In order to know the evolution of a loop in the 2+1D space we have to give the velocity ${\bm V}$ of the loop which contains beside the spatial components also directional velocity which represents rotations of line segments. Since the spatial velocity of a dislocation segment is perpendicular to the spatial line direction $\bm{l}$, the spatial velocity is characterized by a scalar function $v'(\bm{r},\varphi)$. For geometrical reasons the rotation is given by the negative gradient of $v'$ along the line direction $\bm L$\cite{hochrainer2006fundamentals,hochrainer2007three}.  The higher dimensional velocity is thus defined as
\begin{equation}
 \bm V(\bm{r},\varphi)=(v'\sin \varphi,-v' \cos \varphi, -\hat{L}(v')).
\label{eq:veloc_def}
\end{equation}

The time evolution of the system is derived from exterior differential calculus by a Lie derivative in the direction of the generalized velocity $\bm V$, which generalizes the 3D conservation law of $\partial_t \bm \alpha = \nabla \times (\bm{v}\times  \bm{\alpha})$\cite{hochrainer2007three}.
In terms of field $\rho'$ and $k'$ one obtains that 
\begin{eqnarray}
 \partial_t \rho'&=&-\hat{\text{Div}}(\rho' \bm V) + v' k' \rho' \label{eq:rho} \\
 \partial_t k' &=&-v' k'^2-  \hat{L}(\hat{L} (v'))-\hat{V}(k') \label{eq:k}
\end{eqnarray}
with the ``velocity operator'' $\hat{V}$ given by the form
\begin{eqnarray}
 \hat{V}=v' \sin \varphi \, \partial_x - v' \cos \varphi \, \partial_y - \hat{L} (v') \partial_\varphi.
\end{eqnarray}
One can find from Eqs.~(\ref{eq:rho}, \ref{eq:k}) that the time evolution of the quantity $q'=\rho' k'$ is given in the form
\begin{eqnarray}
 \partial_t q'&=&-\hat{\text{Div}}(q' \bm V + \rho'\hat{L}(v'){\bm L}). \label{eq:q}
\end{eqnarray}
(This formula has been first published by Monavari and co-workers \cite{Monavari2016} without derivaton. We provide the derivation in Appendix B).
As it is seen below, the quantity $q'(\bm{r},\varphi)$ is in some sense a more natural quantity to work with than curvature $k'(\bm{r},\varphi)$.

For the further considerations it is useful to give Eq.~(\ref{eq:rho}) in its explicit $\varphi$ dependent form:
\begin{eqnarray}
 \partial_t \rho'&&=-\sin \varphi \, \partial_x (\rho' v') + \cos \varphi \, \partial_y
 (\rho' v') \nonumber \\ && \quad +\partial_\varphi\left\{\cos \varphi \, \partial_x (\rho' v') +\sin \varphi \, \partial_y(\rho' v') \right.  \label{eq:rhot} \\ 
 && \quad \left. +\partial_\varphi\left[\cos \varphi \, \partial_x(q' v')+
 \sin \varphi \, \partial_y (\rho' v')+\partial_\varphi(q' v')
 \right]\right\}
 + v' q'  \nonumber
\end{eqnarray}
in which condition (\ref{eq:noend}) is taken into account. 

In this section we derived kinematic evolution Eqs.~(\ref{eq:rhot}, \ref{eq:q}) for the density fields $\rho'$ and $q'$ defined on the 2+1D space. Together with the velocity field $v'(\bm r, \varphi)$ they form a closed set of kinematic evolution equations. However, due to the large number of degrees of freedom the numerical solution of the resulting equations is not feasible. In the next section we, therefore, continue with reducing the complexity of the problem and develop simplifying assumptions to obtain evolution kinematic equations in the 2D ``real'' space.

\section{Dipole approximation}
As a next step, a dipole approximation is applied for each field appearing in the evolution equations (\ref{eq:rhot},\ref{eq:q}) \cite{wu2018instability}. This means that for the periodic $\varphi$ dependence of the fields we apply a Fourier expansion and we stop at the second terms. With this it is assumed that 
\begin{eqnarray}
 \rho'(\bm{r},\varphi)\approx \rho(\bm{r})+2\cos \varphi \kappa_1(\bm{r})+
 2\sin \varphi \kappa_2(\bm{r}) \label{eq:arho}
\end{eqnarray}
(the reason for the factor 2 is seen below),
\begin{eqnarray}
 v'(\bm{r},\varphi)\approx v^\mathrm m(\bm{r})+\cos \varphi v^\mathrm d_1(\bm{r})+
 \sin \varphi v^\mathrm d_2(\bm{r}), \label{eq:av}
\end{eqnarray}
and
\begin{eqnarray}
 q'(\bm{r},\varphi) \approx q(\bm{r})+\cos \varphi Q_2(\bm{r})-
 \sin \varphi Q_1(\bm{r}). \label{eq:aq}
\end{eqnarray}
Equation \eqref{eq:arho} means that $\rho$ is the net SSD density that is not dependent on the orientation and $\kappa_1$ ($\kappa_2$) is the GND density of dislocations parallel to the $x$ ($y$) axis (angle $\varphi$ is measured from the $x$ axis). Note that this definition coincides with Eqs.~(\ref{eq:kappa1},\ref{eq:kappa2}), hence the same notation. The dipole approximation for the density \eqref{eq:arho}, thus, implies that no directional (screw or edge content) information is available on the SSD content. This is crucial in order to understand, that one may not expect the following theory to exactly specialise to the above sketched theory of straight parallel edge dislocations (where only dislocations parallel to the $y$ axis are present) without explicitly incorporating the available directional information on the SSD. Considering direction information for SSD would require to use at least two more Fourier coefficients.

\begin{figure}[!ht]
\includegraphics[angle=0,width=7.5cm]{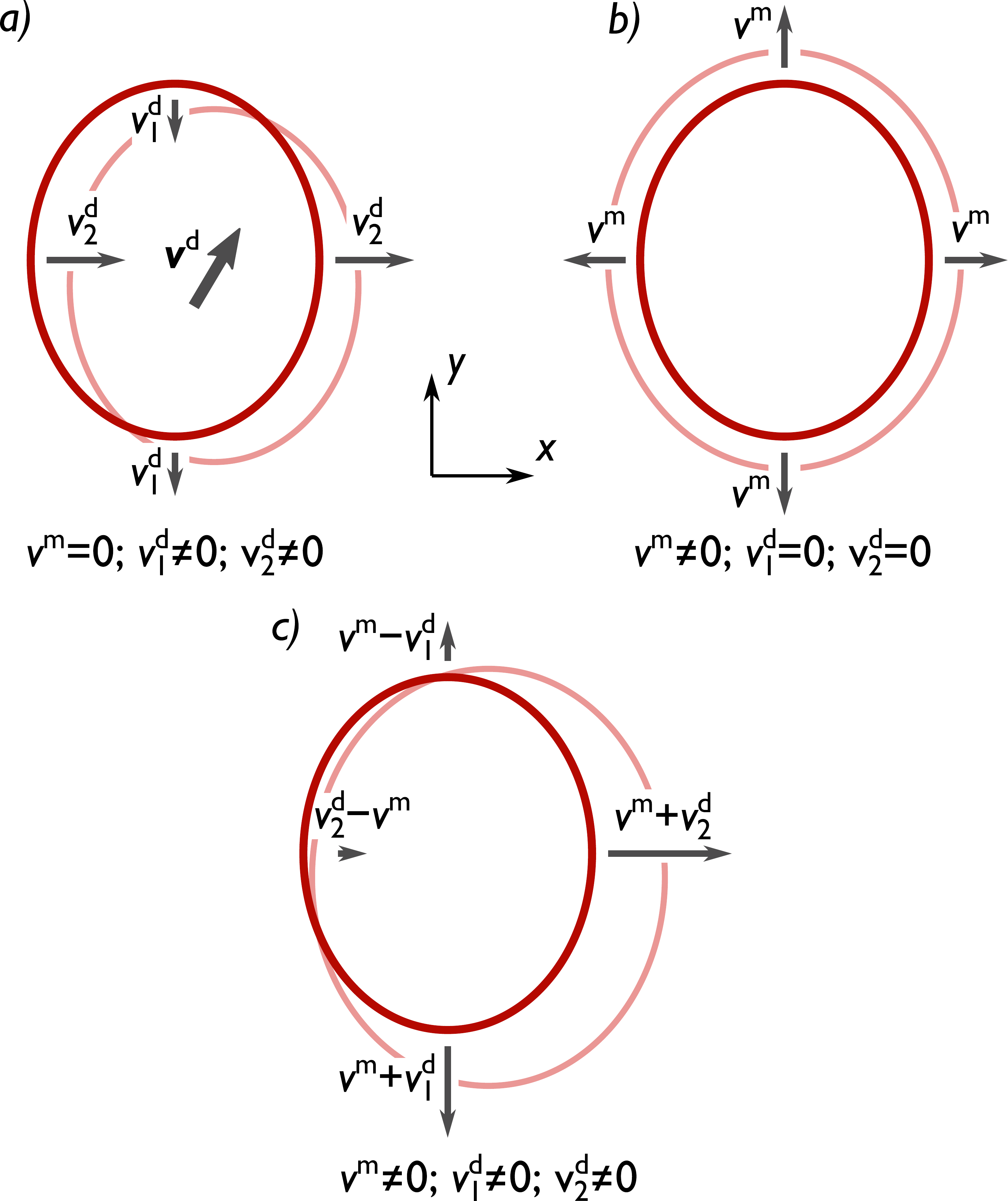}
\caption{Sketch of the meaning of velocities $v^\mathrm m$ and $v^\mathrm d_{1,2}$ defined by Eq.~\eqref{eq:av} in the case of a dislocation loop in the $xy$ plane. Here, for simplicity, it is assumed that the velocities are constant in space. The dark red loop represents the original loop and the one in light red is the shape obtained after a short time. Note, that the velocities along the loop change continuously according to Eq.~\eqref{eq:av}, the arrows represent only the velocities of the segments being parallel to $x$ and $y$ axes. (a): When $v^\mathrm m = 0$ but $v^\mathrm d_1 \ne 0$ and $v^\mathrm d_2 \ne 0$, then the loop moves as a ``rigid body'' (as in the 2D case, see Fig.~\ref{fig:velocities}(a)). Direction $\bm v^\mathrm d$ is also noted: as explained in the text, part of the loop parallel to $\bm v^\mathrm d$ has zero velocity, and the part being perpendicular to $\bm v^\mathrm d$ has the largest speed. (b): If $v^\mathrm d_1 = v^\mathrm d_2 = 0$ then the loop expands with velocity $v^\mathrm m$. Here the shape of the loop changes, analogously to the situation in 2D, see Fig.~\ref{fig:velocities}(b). Note that the velocities in the $x$ and $y$ direction are equal due to the assumption of Eq.~\eqref{eq:av}. (c): Superposition of panels a) and b) when all velocity components $v^\mathrm m$ and $v^\mathrm d_{1,2}$ are non-zero. In this case translation and expansion of the loop takes place simultaneously.}
\label{fig:velocities_3d}
\end{figure}

The dipole approximation for the velocity \eqref{eq:av} introduces three scalar velocity terms: $v^\mathrm m$, $v^\mathrm d_1$, and $v^\mathrm d_2$. According to the definition of the velocity $v'$ [Eq.~\eqref{eq:veloc_def}], the meaning of these terms are visualized in Fig.~\ref{fig:velocities_3d} in case of a single dislocation loop. It is clear that $v^\mathrm m$ is a \emph{mean} velocity  of dislocations of all characters, whereas the \emph{drift} velocity vector $\bm{v}^\mathrm d = (v^\mathrm d_1, v^\mathrm d_2)$ characterises the direction of dislocations ($\bm{l} \parallel \bm{v}^\mathrm d $)  for which the opposite characters show the largest velocity difference given by the twice the modulus $v^\mathrm d =2 | \bm{v}^\mathrm d| $. It is important to not confuse the drift velocity vector with any kind of dislocation velocity vector which is always perpendicular to the local line direction. The comparison of Figs.~\ref{fig:velocities} and \ref{fig:velocities_3d} clearly demonstrates the analogy between the $v^\mathrm m$ and $v^\mathrm d$ velocities for the different models, and explains the identical notation.

Regarding the dipole approximation \eqref{eq:aq} for the curvature density $q'= \rho' k'$ we note that $\rho'$ and $q'$ are connected by the solenoidality requirement \eqref{eq:noend}. The curvature difference vector $\bm{Q}$, which indicates the direction of dislocations with maximum difference in curvature between dislocations of opposite orientation, may therefore not be independently defined from the density approximation. While the dipole approximations will violate solenoidality on the higher dimensional space, we require that the dipole approximation of the divergence on the higher dimensional space, i.e., of the quantity
\begin{equation}
 G(\bm{r},\varphi):=\cos \varphi \, \partial_x \rho'(\bm{r},\varphi)+\sin \varphi \, \partial_y \rho'(\bm{r},\varphi)+\partial_\varphi q'(\bm{r},\varphi),
\end{equation}
has to vanish. By taking the integral 
\begin{eqnarray}
 G^\mathrm m(\bm{r})=\frac{1}{2\pi}\int_0^{2\pi}G(\bm{r},\varphi) \mathrm d\varphi,
\end{eqnarray}
one can see from (\ref{eq:arho}) that the vector $\bm{\kappa}$ needs to be solenoidaly,
\begin{eqnarray}
 G^\mathrm m(\bm{r})=\partial_x \kappa_1(\bm{r})+\partial_y\kappa_2(\bm{r})=0. \label{eq:ckappa}
\end{eqnarray}
This is an important ``constraint'' relation between the $\kappa_{1,2}$ fields. It expresses the fact that dislocation loops cannot end in the bulk of a sample \cite{hochrainer2007three,hochrainer2015multipole}. 

Also the Fourier coefficients
\begin{eqnarray}
 G^\mathrm d_1(\bm{r})=\frac{1}{2\pi}\int_0^{2\pi} \cos \varphi \, G(\bm{r},\varphi) \mathrm d\varphi
\end{eqnarray}
and
\begin{eqnarray}
 G^\mathrm d_2(\bm{r})=\frac{1}{2\pi}\int_0^{2\pi} \sin \varphi \, G(\bm{r},\varphi) \mathrm d\varphi
\end{eqnarray}
are supposed to vanish, $G^\mathrm d_{1,2} = 0$. With the quantities introduced in Eqs.~(\ref{eq:arho}, \ref{eq:aq}) this leads to the consistency requirements
\begin{eqnarray}
 Q_1(\bm{r})=\partial_x \rho(\bm{r}), \label{eq:crho1}
\end{eqnarray}
and
\begin{eqnarray}
 Q_2(\bm{r})=\partial_y \rho(\bm{r}). \label{eq:crho2}
\end{eqnarray}
Note, that unlike Eq.~(\ref{eq:ckappa}), the relation of $ \bm{Q}$ to the density variables, i.e., Eqs.~(\ref{eq:crho1},\ref{eq:crho2}) depends on the number of considered Fourier coefficients. Terms related to next order Fourier coefficients are neglected (for details see Hochrainer \cite{hochrainer2015multipole}).

Next we shall determine the evolution equations of the fields $\rho$, $\kappa_{1,2}$, and $q$. By substituting expressions (\ref{eq:arho},\ref{eq:av},\ref{eq:aq}) into
Eq.~(\ref{eq:rhot}) we  arrive at the evolution equation:
\begin{eqnarray}
 \frac{1}{2\pi}\int_0^{2\pi} \partial_t \rho' \mathrm d\varphi=\partial_t \rho
\end{eqnarray}
with
\begin{eqnarray}
 \partial_t \rho&=&-\frac{1}{2}\partial_x(\rho v^\mathrm d_2)+\frac{1}{2}\partial_y(\rho v^\mathrm d_1)+\partial_y(\kappa_1 v^\mathrm m)-\partial_x(\kappa_2 v^\mathrm m)
 \nonumber \\ &&+\frac{1}{2}q v^\mathrm m+\frac{1}{2}v^\mathrm d_1 Q_2-v^\mathrm d_2 Q_1.
 \label{eq:trho1}
\end{eqnarray}
By substituting the relations (\ref{eq:crho1},\ref{eq:crho2}) into the above equations we obtain
\begin{eqnarray}
 \partial_t \rho&=&-\frac{1}{2}\partial_x(\rho v^\mathrm d_2)+\frac{1}{2}\partial_y(\rho v^\mathrm d_1)+\partial_y(\kappa_1 v^\mathrm m)-\partial_x(\kappa_2 v^\mathrm m)
 \nonumber \\ &&+qv^\mathrm m+ \frac{1}{2}v^\mathrm d_1\partial_y \rho-\frac{1}{2}v^\mathrm d_2 \partial_x \rho.
 \label{eq:trho0}
\end{eqnarray}
where we also neglected terms related to higher order Fourier coefficients. It is useful to rewrite the above equation into the form 
\begin{eqnarray}
 \partial_t \rho&=&-\partial_x(\rho v^\mathrm d_2)+\partial_y(\rho v^\mathrm d_1)+\partial_y(\kappa_1 v^\mathrm m)-\partial_x(\kappa_2 v^\mathrm m)
 \nonumber \\ &&+qv^\mathrm m+ \frac{1}{2}\rho\partial_y v^\mathrm d_1- \frac{1}{2}\rho\partial_x v^\mathrm d_2.
 \label{eq:trho2}
\end{eqnarray}
We have now arrived at the kinematic evolution equation of $\rho$ in the ``dipole'' approximation. The comparison of this equation with that of the 2D case [Eq.~\eqref{eq:rr}] shows remarkable similarity. It should be noted, however, that the last two terms are not ``compatible'' with the 2D results, if the current equations would be simply adopted for straight edge dislocation case. (The later would mean that one assumes $\bm b$ to be parallel with the $x$ axis and takes $\kappa_1 = 0$ and $q=0$.) As discussed above, this is the result of the closing approximation we indirectly apply with the forms given by Eqs.~(\ref{eq:arho},\ref{eq:av},\ref{eq:aq})\cite{hochrainer2015multipole}. One cannot expect that it is directly applicable for straight dislocations. 
In order to allow for incorporating this information, multipliers $\lambda_1,\lambda_2$ are introduced. For straight dislocations $\lambda_{1,2}=0$ and in the dipole approximation applied here $\lambda_{1,2}=1/2$. So,
\begin{eqnarray}
 \partial_t \rho&=&-\partial_x(\rho v^\mathrm d_2)+\partial_y(\rho v^\mathrm d_1)+\partial_y(\kappa_1 v^\mathrm m)-\partial_x(\kappa_2 v^\mathrm m)
 \nonumber \\ &&+qv^\mathrm m+ \lambda_1\rho\partial_y v^\mathrm d_1- \lambda_2\rho\partial_x v^\mathrm d_2.
 \label{eq:trho}
\end{eqnarray}

For the further considerations we make two important observations:
\begin{eqnarray}
 &&\cos \varphi \left(\hat{\text{Div}}(\rho' {\bm V}) + q'v'\right) \nonumber \\ 
 &&\quad =\partial_y(\rho' v')-\partial_\varphi\left(-q'v'\sin \varphi-\rho'\hat{L}(v') \cos \varphi\right), \label{eq:cosrho}
\end{eqnarray}
and
\begin{eqnarray}
 &&\sin \varphi \left(\hat{\text{Div}}(\rho' {\bm V})+q'v'\right) \nonumber \\
 &&\quad =-\partial_x(\rho' v')+\partial_\varphi\left(q'v'\cos \varphi+\rho'\hat{L}(v') \sin \varphi\right). \label{eq:sinrho}
\end{eqnarray}
The detailed derivation of the above identities is explained in Appendix A.
By taking 
\begin{eqnarray}
 \frac{1}{2\pi}\int_0^{2\pi} \cos \varphi \, \partial_t \rho' \mathrm d\varphi=\partial_t \kappa_1
\end{eqnarray}
with Eqs.~(\ref{eq:arho},\ref{eq:av},\ref{eq:cosrho}) one finds that
\begin{eqnarray}
 \partial_t \kappa_1&=&\partial_y(\rho v^\mathrm m+\kappa_1 v^\mathrm d_1+\kappa_2 v^\mathrm d_2). \label{eq:tkappa1}
\end{eqnarray}
In a similar way by calculating
\begin{eqnarray}
 \frac{1}{2\pi}\int_0^{2\pi} \sin \varphi \, \partial_t \rho' \mathrm d\varphi=\partial_t \kappa_2
\end{eqnarray}
one arrives at
\begin{eqnarray}
 \partial_t \kappa_2&=&-\partial_x(\rho v^\mathrm m+\kappa_1 v^\mathrm d_1+\kappa_2 v^\mathrm d_2). \label{eq:tkappa2}
\end{eqnarray}
Again, comparison with Eq.~\eqref{eq:kk} shows a clear analogy with the 2D case, here without any additional terms due to the ``dipole'' approximation.

For the evolution equation of the curvature $q$ we use Eq.~(\ref{eq:q}) in its explicit form (see  Hochrainer\cite{hochrainer2015multipole}.)
\begin{equation}
\begin{split}
 \partial_t q=&-\partial_x \frac{1}{2\pi}\int_0^{2\pi}\left[q' v' \sin \varphi+\rho' \hat{L}(v) \cos \varphi 
 \right]\mathrm d\varphi \\
 &-\partial_y \frac{1}{2\pi}\int_0^{2\pi}\left[-q' v' \cos \varphi+\rho' \hat{L}(v) \sin \varphi 
 \right]\mathrm d\varphi
\end{split}
\end{equation}
indicating the important fact that $q$ is a conserved quantity. Again in the ``dipole'' approximation  one obtains
\begin{widetext}
\begin{eqnarray}
 \partial_t q&=&\partial_x\left[-qv^\mathrm d_2+v^\mathrm mQ_1-\frac{1}{2}\partial_x\left(\rho v^\mathrm m+\frac{1}{2}\kappa_2 v^\mathrm d_2+\frac{3}{2}\kappa_1v^\mathrm d_1\right)-\frac{1}{4}\partial_y\left(\kappa_1 v^\mathrm d_2+\kappa_2 v^\mathrm d_1\right) 
 \right] \nonumber \\
 &+&\partial_y\left[+qv^\mathrm d_1+v^\mathrm mQ_2-\frac{1}{2}\partial_y\left(\rho v^\mathrm m+\frac{1}{2}\kappa_1 v^\mathrm d_1+\frac{3}{2}\kappa_2v^\mathrm d_2\right)-\frac{1}{4}\partial_x\left(\kappa_2 v^\mathrm d_2+\kappa_1 v^\mathrm d_1\right) 
 \right]. 
 \label{eq:tk0}
\end{eqnarray}
\end{widetext}
If we assume a nearly homogeneous system and neglect all terms in Eq.~(\ref{eq:tk0}) which contain second derivatives, we arrive at the simplified form
\begin{eqnarray}
 \partial_t q&=&-\partial_x\left(qv^\mathrm d_2-v^\mathrm m Q_1\right) 
 +\partial_y\left(q v^\mathrm d_1+v^\mathrm m Q_2\right),
 \label{eq:tk}
\end{eqnarray}
which we will use subsequently, however, the general case can be treated in a similar way. 

It should be noted that the ``truncation'' procedure applied above corresponds to a ``natural'' closing approximation. One may consider higher order terms in the Fourier expansion of the fields $\rho'(\bm{r},\varphi)$,  
$q'(\bm{r},\varphi)$, and $v'(\bm{r},\varphi)$ and apply some other closer approximations (see T. Hochrainer and M.~Monavari \emph{et al.}~for details \cite{hochrainer2015multipole,Monavari2016}).  Nevertheless, the general structure of the  evolution equations remains the same. 

To summarize this section, we applied a ``dipole'' type approximation to the kinematic equations in 2+1D. The name refers to the fact that the $\varphi$-dependent terms were expanded up to the first  order Fourier coefficients. We emphasize, that no other assumptions on the microstucture (e.g., the presence of dislocation dipoles) were made. Using the approximation we derived kinematic evolution equations (\ref{eq:trho},\ref{eq:tkappa1},\ref{eq:tkappa2},\ref{eq:tk}) that are formulated in terms of the variables $\rho$, $\kappa_{1,2}$, $q$, $v^\mathrm m$, $v^\mathrm d_{1,2}$ (note that $Q_{1,2}$ derive from $ \rho$ by Eqs.\ (\ref{eq:crho1},\ref{eq:crho2})) defined in the ``real'' 2D space. 
The approximation was motivated by the fact that the resulting evolution equations are analogous to the 2D case described in Sec.~\ref{sec:2d} which was found to represent dislocation dynamics in sufficient detail. 

\section{Plastic distortion}

Before proceeding to the dynamics of the system we have to discuss what is the plastic distortion tensor $\bm \beta^\mathrm p$ and its rate $\dot{\bm \beta}^\mathrm p$ for the geometry considered. As it is known the dislocation density tensor $\alpha_{ij}=e_{ikl}\partial_k \beta^\mathrm p_{lj}$ where $e_{ikl}$ is the Levi-Civita  tensor. As a consequence, for the general case, $\bm \alpha$ does not uniquely determine $\bm \beta^\mathrm p$. For the problem considered here, however, this is not the case. Since we assumed that dislocation loops can evolve only in their slip planes only the $i=3$ components of $\beta^\mathrm p_{ij}$ are different from zero \cite{kroner1981continuum}, and $\beta^\mathrm p_{3j}(\bm{r})=\beta_0(\bm{r}) b_j$ where $\beta_0(\bm{r})$ is a scalar that can be calculated from $\bm \alpha$ as follows: from the definition of $\kappa_1$ and $\kappa_2$ given by Eq.~(\ref{eq:arho}) one can see that
\begin{eqnarray}
 \kappa_1&=&\partial_y \beta_0 \label{eq:k1}, \\
 \kappa_2&=&-\partial_x \beta_0. \label{eq:k2} 
\end{eqnarray}
Now, let us introduce the vector $\bm{F}=(-\kappa_2,\kappa_1,0)$. Assuming that we are in a given slip plane, (i.e., $z$ is fixed), one can find that
due to the condition $\partial_x\kappa_1+\partial_y \kappa_2=0$ $\bm{F}$ is $\text{Curl}$-free ($\text{Curl}\,  \bm F = 0$). As a consequence $\bm{F}$ can be given as the gradient of a scalar field. It is straightforward to see that $F_i=\partial_i \beta_0$ with $z$ coordinate considered as a fixed parameter. With this, $\beta_0$ can be calculated from the GND density with the integral
\begin{eqnarray}
 \beta_0(\bm{r})=\int_\infty^{\bm{r}} F_i \, \mathrm dr_i \label{eq:beta}
\end{eqnarray}
 where the integration can be carried out for any curve that is in a plane perpendicular to the $z$ direction. (Here we assumed that all fields go to zero at infinity.)

Concerning the rate of plastic deformation $\dot{\bm \beta^\mathrm p}$ it is obviously determined by $\dot{\beta_0}$ as $\dot\beta^\mathrm p_{3j}(\bm{r})=\dot\beta_0(\bm{r}) b_j$ (the other components are zero). From Eqs.~(\ref{eq:k1},\ref{eq:k2},\ref{eq:beta}) one gets
\begin{eqnarray}
 \dot{\beta}_0(\bm{r})=\int_\infty^{\bm{r}} \partial_t{F}_i \mathrm dr_i \label{eq:dotbeta}
\end{eqnarray}
with $\partial_t\bm{F}=(-\partial_t\kappa_2,\partial_t\kappa_1,0)$. 
From Eqs.~(\ref{eq:tkappa1},\ref{eq:tkappa2})
\begin{eqnarray}
\partial_t F_1&=&\partial_x(\rho v^\mathrm m+\kappa_1 v^\mathrm d_1+\kappa_2 v^\mathrm d_2) \\
\partial_t F_2&=&\partial_y(\rho v^\mathrm m+\kappa_1 v^\mathrm d_1+\kappa_2 v^\mathrm d_2)
\end{eqnarray}
leading to
\begin{eqnarray}
 \dot{\beta}_0=\rho v^\mathrm m+\kappa_1 v^\mathrm d_1+\kappa_2 v^\mathrm d_2. \label{eq:betadot}
\end{eqnarray}
The analogy with the 2D model is again fulfilled [cf.~Eq.~\eqref{eq:gg}].

\section{The evolution of the plastic potential}

So far we have derived kinematic evolution equations for the curved dislocation system in the frame of a dipole approximation. The resulting Eqs.~(\ref{eq:trho},\ref{eq:tkappa1},\ref{eq:tkappa2},\ref{eq:tk}) do not specify how to obtain velocities $v^\mathrm m$ and $v^\mathrm d_{1,2}$ from the state variables $\rho$, $\kappa_{1,2}$ and $q$. In order to perform this step and to arrive at the desired closed set of dynamic equations we generalize the thermodynamic considerations of the 2D model described in Sec.~\ref{sec:2d}. As a starting point we assume that there is a scalar functional of the fields $\rho$, $\kappa_{1,2}$ and $q$ that cannot increase during the evolution of the system. We recall that in the 2D case this functional was derived from microscopic considerations \cite{groma2016dislocation}, which, due to the complexity of the microstructure, is not feasible for the present 3D case. It should be noted again, as it is explained in Appendix D, that this functional is different from the coarse grained elastic energy.

For the curved dislocation problem, at this stage, we do not specify the actual form of the functional (for that, see Sec.~\ref{sec:pp}), rather, in analogy with Sec.~\ref{sec:2d}, first we are going to obtain mobility laws that are, on the one hand, consistent with the 2D case and, on the other hand, guarantee that this functional cannot increase during the evolution of the system. The functional is denoted by $P[\rho,\kappa_1,\kappa_2,q]$ and it is  called ``plastic potential''.

In general its time derivative is given by the equation
\begin{eqnarray}
 \dot P&=&\int \left[\frac{\delta P}{\delta \rho} \partial_t \rho+
 \frac{\delta P}{\delta \kappa_1} \partial_t \kappa_1 \right .\nonumber \\&&\left .+
 \frac{\delta P}{\delta \kappa_2} \partial_t \kappa_2+ 
 \frac{\delta P}{\delta q} \partial_t q
 \right] \mathrm d V,  \label{eq:Pt}
\end{eqnarray}
where  $\delta P/\delta X$ denotes the functional derivative of $P$ with respect to the field $X$. For shorter notations we introduce the quantities 
\begin{equation}
 \mu_\rho=\frac{\delta P}{\delta \rho},\ \mu_{\kappa_1}=\frac{\delta P}{\delta \kappa_1},\  \mu_{\kappa_2}=\frac{\delta P}{\delta \kappa_2},\ \mu_{q}=\frac{\delta P}{\delta q}.
\end{equation}
They could be called  the appropriate ``chemical potentials''. The name comes from the formal analogy with chemical potential used in thermodynamics. These quantities are in fact intensive state variables, we note, however, that due to the friction-like mobility laws introduced below these quantities may not be constant in equilibrium.

By substituting the kinematic Eqs.~(\ref{eq:trho},\ref{eq:tkappa1},\ref{eq:tkappa2},\ref{eq:tk}) into Eq.~(\ref{eq:Pt}) and then performing partial integrations one arrives at the inequality
\begin{equation}
\begin{split}
 \dot P=&- \int b\rho \left[ (\tau^*
 +\tau^\mathrm m) v^\mathrm m+ \left(\frac{\kappa_1}{\rho} \tau^*+\tau^\mathrm d_1\right) v^\mathrm d_1 \right. \\
 &+ \left. \left(\frac{\kappa_2}{\rho} \tau^*+\tau^\mathrm d_2\right) v^\mathrm d_2 
 \right] \mathrm d V \leqq 0,
\end{split}
\label{eq:pp}
\end{equation}
where the following terms with stress dimension were introduced
\begin{eqnarray}
 \tau^*&=& \frac1b \left[(\partial_y \mu_{\kappa_1})-(\partial_x \mu_{\kappa_2}) \right], \label{eq:taustar} \\
 \tau_1^\mathrm d &=& \frac 1{b \rho} \left[ (\partial_y \mu_{\rho})\rho+(\partial_y\mu_q)q +\partial_y(\lambda_1 \mu_\rho \rho) \right], \label{eq:tau1d} \\
 \tau_2^\mathrm d &=& -\frac 1{b \rho} \left[ (\partial_x \mu_{\rho})\rho+(\partial_x \mu_q)q + \partial_x(\lambda_2\mu_\rho \rho) \right], \label{eq:tau2d} \\
 \tau^\mathrm m &=& \frac 1{b \rho} \left[ (\partial_y \mu_\rho)\kappa_1-(\partial_x \mu_\rho)\kappa_2 \right. \label{eq:taum} \\ && \left. + (\partial_x \mu_q) Q_1 + (\partial_y \mu_q) Q_2 -\mu_\rho q \right]. \nonumber
\end{eqnarray}
For these terms notations $\tau^*$ and $\tau^\mathrm d_{1,2}$ were used based on the analogy with the 2D case [cf.~Eqs.~(\ref{eq:tau_star},\ref{eq:tau_d})], whereas $\tau^\mathrm m$ does not have a corresponding term in the 2D model. 

To guarantee that the plastic potential does not increase in time the intergrand of Eq.~(\ref{eq:pp}) must everywhere be non-negative, that is,
\begin{eqnarray}
 &&(\tau^*+\tau^\mathrm m)v^\mathrm m+\left(\frac{\kappa_1}{\rho} \tau^*+\tau^\mathrm d_1\right)v^\mathrm d_1
 \nonumber \\ && \quad +
 \left(\frac{\kappa_2}{\rho} \tau^*+\tau^\mathrm d_2\right)v^\mathrm d_2\geqq 0 \label{eq:ine}
\end{eqnarray}

Next, we now introduce the mobility laws for curved dislocations that are analogous to the 2D case [Eqs.~(\ref{eq:v0},\ref{eq:v1})] and, thus, read as
\begin{eqnarray}
 v^\mathrm m&=&M_0 b\chi (\tau^*), \label{eq:v0_3d} \\
 v^\mathrm d_{1,2}&=&M_0b\left( \frac{\kappa_{1,2}}{\rho} (\tau^*-\chi(\tau^*)) +\tau^\mathrm d_{1,2}\right). \label{eq:v1_3d}
\end{eqnarray}
This means, that up to the local yield stress $\tau^\mathrm y$ the mean velocity $v^\mathrm m$ is zero, but $v^\mathrm d_{1,2}$ may have a non-zero value, the situation sketched in Fig.~\ref{fig:velocities_3d}(a). Above the yield stress, however, all the velocity terms may differ from zero, which corresponds to Fig.~\ref{fig:velocities_3d}(c). The exact value of $\tau^\mathrm y$ will be defined later.

The question we address in the following is whether these mobility laws are consistent with the inequality (\ref{eq:ine}). It is easy to see by simple substitution that the condition is fulfilled if $|\tau^*| \leqq \tau^\mathrm y$ (non-flowing regime). When $|\tau^*| > \tau^\mathrm y$ (flowing regime) the inequality (\ref{eq:ine}) reads as
\begin{equation}
\begin{split}
 & \left[\tau^* -s\tau^\mathrm y+\left(1-\frac{\kappa_1^2}{\rho^2}-\frac{\kappa_2^2}{\rho^2}\right)s\tau^\mathrm y \right. \\
 &\left .+\frac{\kappa_1}{\rho}B^\mathrm d_1+\frac{\kappa_2}{\rho}B^\mathrm d_2+B^\mathrm m\right](\tau^* -s\tau^\mathrm y) \\
 &+ \left(\frac{\kappa_1}{\rho}(\tau^* -s\tau^\mathrm y)+B^\mathrm d_1\right) B^\mathrm d_1 \\
 &+ \left(\frac{\kappa_2}{\rho}(\tau^* -s\tau^\mathrm y)+B^\mathrm d_2\right)
 B^\mathrm d_2\geqq 0,
 \end{split}
 \label{eq:bieq}
\end{equation}
where $s=\sign(\tau^*)$, and the following auxiliary stress terms are introduced
\begin{eqnarray}
 B^\mathrm d_1&=&  s\frac{\kappa_1}{\rho} \tau^\mathrm y+\tau^\mathrm d_1, \\
 B^\mathrm d_2&=&  s\frac{\kappa_2}{\rho} \tau^\mathrm y+\tau^\mathrm d_2, \\
 B^\mathrm m&=& \tau^\mathrm m - \frac{\kappa_{1}}{\rho} \tau^\mathrm d_1 - \frac{\kappa_{2}}{\rho} \tau^\mathrm d_2. \label{eq:b0} 
\end{eqnarray}
Notice that $B^\mathrm d_1 \cdot M_0 b$ and $B^\mathrm d_2 \cdot M_0 b$ are the velocities $v^\mathrm d_1$ and $v^\mathrm d_2$ at $|\tau^*| \geqq \tau^\mathrm y$, respectively. The inequality (\ref{eq:bieq}) can now be rewritten as 
\begin{equation}
\begin{split}
 & \left(1-\frac{\kappa_1^2}{\rho^2}-\frac{\kappa_2^2}{\rho^2}\right)(\tau^* -s\tau^\mathrm y)^2 \\&+
\begin{pmatrix}
       \tau^* -s\tau^\mathrm y, &
       B^\mathrm d_1
\end{pmatrix}
\begin{pmatrix}
        \frac{\kappa_1^2}{\rho^2}  & \frac{\kappa_1}{\rho} \\
        \frac{\kappa_1}{\rho} & 1
\end{pmatrix}
\begin{pmatrix}
       \tau^* -s\tau^\mathrm y\\
       B^\mathrm d_1
\end{pmatrix} \\&+
\begin{pmatrix}
       \tau^*-s\tau^\mathrm y, &
       B^\mathrm d_2
\end{pmatrix}
\begin{pmatrix}
        \frac{\kappa_2^2}{\rho^2}  & \frac{\kappa_2}{\rho} \\
        \frac{\kappa_2}{\rho} & 1
\end{pmatrix}
\begin{pmatrix}
       \tau^* -s\tau^\mathrm y\\
       B^\mathrm d_2
\end{pmatrix} \\
& +\left[ B^\mathrm m + \left(\frac{\kappa_1^2}{\rho^2}+\frac{\kappa_2^2}{\rho^2} - 1\right)s\tau^\mathrm y \right](\tau^*-s\tau^\mathrm y)\geqq 0.
\end{split}
\end{equation}
Combining the first and the fourth term and realizing that the second and third term are always non-negative, unconditional non-negativity requires that 
\begin{eqnarray}
	\left[ B^\mathrm m+\left(1-\frac{\kappa_1^2}{\rho^2}-\frac{\kappa_2^2}{\rho^2}\right)\tau^*
	\right](\tau^* - s\tau^\mathrm y)\geqq 0.
	\label{eq:bb}
\end{eqnarray}
With the introduction of the dimensionless quantity
\begin{eqnarray}
 \beta=\frac{B^\mathrm m}{\mu b \sqrt{\rho}\left(1-\frac{\kappa_1^2}{\rho^2}-\frac{\kappa_2^2}{\rho^2}\right)}.
\label{eq:betad}
\end{eqnarray}
the inequality reads as
\begin{eqnarray}
	(\tau^* + \beta \mu b \sqrt{\rho})(\tau^* -s\tau^\mathrm y)\geqq 0.
	\label{eq:bbb}
\end{eqnarray}
Note that the non-negativity of $\rho'$ in Eq.~\eqref{eq:arho} implies that the term $\left( 1-\frac{\kappa_1^2}{\rho^2}-\frac{\kappa_2^2}{\rho^2} \right)$ in the denominator of Eq.~\eqref{eq:betad} is always positive.   

According to the well-known Taylor hardening law the yield stress is
\begin{eqnarray}
 \tau^\mathrm y=\alpha \mu b \sqrt{\rho}.
\end{eqnarray}
Condition Eq.~(\ref{eq:bbb}) is consequently fulfilled if $|\beta|<\alpha$. One can see from Eq.~(\ref{eq:b0}) that $\beta$ contains a term that is proportional to $q/\rho^{3/2}$ which is the ratio of the average dislocation spacing $ 1/\sqrt{\rho} $ and the average radius of curvature $\rho/q$. For common dislocation configurations this ratio is supposed to be small. The other terms in $\beta$ are proportional to spatial derivatives of the various fields. So for nearly homogeneous configurations at small $q/\rho^{3/2}$ ratio $|\beta|$ is assumed to be small compare to the Taylor coefficient $\alpha$, which usually takes values between $0.3$ and $0.4$ \cite{kocks_m03}. On the other hand, in highly inhomogeneous or strongly curved dislocation configurations, considering the named global Taylor coefficient $\alpha$ becomes questionable on a local level. In this case, we therefore propose to ensure condition Eq.~(\ref{eq:bbb}) by making $\alpha$ a function of $\beta$ and thus of the local dislocation state. As the simplest possible $\alpha(\beta)$ function one may take
\begin{eqnarray}
 \alpha(\beta)=\left \{
 \begin{array}{ll}
      \alpha_0 \ \ & {\rm if} \ \ |\beta|<|\alpha_0| \\
      |\beta|  & {\rm if} \ \ |\beta|\geq|\alpha_0|,
    \end{array}
 \right.
\end{eqnarray}
where now $\alpha_0$ denotes the constant global coefficient. 

The finding that $\alpha$ becomes a function of the local dislocation state is consistent with the general experimental observation that $\alpha$ is (weakly) dependent on the type of dislocation pattern developing upon different modes and levels of deformation \cite{Basinski_b79}. However, deriving the state dependence of $\alpha$ will require further investigations.

\section{The proposed form of the plastic potential}
\label{sec:pp}

The last step to arrive at dynamic evolution equations for curved dislocations is to specify the form of the plastic potential $P$. The stress terms $\tau^*$ and $\tau^\mathrm d_{1,2}$ then follow from Eqs.~(\ref{eq:taustar}-\ref{eq:tau2d}). It is important however to distinguish the so-called \emph{mean-field} stress tensor $\bm \sigma$ and the resolved shear stress $\tau^\mathrm{mf} = (1/b)\bm n \cdot \bm \sigma \cdot \bm b$ in the glide plane ($\bm n$ being the glide plane normal vector) that are due to the long-range stress field generated by the GND dislocations and the boundary tractions and displacements. The reason is that this is a measurable quantity (average local stress) and boundary conditions can only be formulated in terms of $\bm \sigma$. In order to do so here we again follow the route developed for 2D straight dislocations (see Groma \cite{groma2016dislocation}) and split the plastic potential into a ``mean field'' and a ``correlation'' part
\begin{equation}
    P=P^\mathrm{mf}+P^\mathrm{corr}. \label{eq:P_split}
\end{equation}

The first term $P^\mathrm{mf}$ is the mean field elastic energy of the system, which means, that if the dislocations were distributed randomly, then $P^\mathrm{mf}$ would exactly be the elastic energy. It reads as\cite{groma2010variational,groma2016dislocation,groma2019statistical}
\begin{equation}
\begin{split}
 & P^\mathrm{mf}[\bm \chi, \kappa_1, \kappa_2] \\
 &\quad =\int \left[
 -\frac{1}{2} (\text{Inc}\, \bm\chi)_{ij} C^{-1}_{jikl} (\text{Inc}\, \bm\chi)_{lk}+\chi_{ij}\eta_{ji}\right]\mathrm dV,
\end{split}
\end{equation}
where $\text{Inc}$ is the incompatibility operator, $\bm \chi$ is the stress potential tensor, $\bm C$ is the elastic modulus tensor, and $\bm \eta$ is the incompatibility tensor related to the dislocation density tensor as \cite{kroner1981continuum,groma2019statistical}
\begin{eqnarray}
 \eta_{ij}=\frac{1}{2}(e_{iln}\partial_n \alpha_{jl}+e_{jln}\partial_n \alpha_{il}).
\end{eqnarray}
This means that the dependence of $P^\mathrm{mf}$ on $\kappa_1$ and $\kappa_2$ appears implicitly through the dislocation density tensor via Eq.~(\ref{eq: def alpha}) and the $i=3$ and $j=3$ components of $\alpha_{ij}$ are zero. 

For $P^\mathrm{mf}$ the auxiliary variable $\bm \chi$ was introduced. As it was shown by Groma \emph{et al.}\cite{groma2010variational}\ the stress equilibrium equation can then be obtained by
\begin{eqnarray}
 \frac{\delta P^\mathrm{mf}}{\delta \chi_{ij}}=0,
 \label{eq:equilibrium}
\end{eqnarray}
an the stress tensor follows from its solution as
\begin{eqnarray}
 \bm \sigma=\text{Inc} \, \bm \chi.
\end{eqnarray}
The equilibrium Eq.~(\ref{eq:equilibrium}) lets us to introduce the boundary conditions for surface tractions and displacements for a given sample geometry. As it is shown in Appendix \ref{sec:mean_field}, the mean field resolved shear stress $\tau^\mathrm{mf}$ follows as
\begin{equation}
    \tau^\mathrm{mf}= \frac1b \left[ \left(\partial_y \frac{\delta P^\mathrm{mf}}{\delta \kappa_1} \right) - \left(\partial_x \frac{\delta P^\mathrm{mf}}{\delta \kappa_2} \right) \right].
\end{equation}
According to Eq.~(\ref{eq:taustar}) if $P^\mathrm{corr}$ was zero (that is, we assumed a random distribution of dislocations) then $\tau^*$ would be equal to $\tau^\mathrm{mf}$.

The ``correlation'' part $P^\mathrm{corr}$ represents a correction to the plastic potential due to the fact that dislocations are not positioned randomly but develop spatial correlations. This part cannot depend on $\bm \chi$ otherwise the stress equilibrium Eq.~(\ref{eq:equilibrium}) would be violated. Its simplest possible form can be obtained from symmetry and dimensionality arguments and, based on the 2D straight parallel dislocation problem, reads as
\begin{equation}
\begin{split}
 & P^\mathrm{corr}[\rho, \kappa_1, \kappa_2, q] \\ &\quad =\int G b^2 \left[A \rho \ln\left(\frac{\rho}{\rho_0}\right)+ 
 \frac{\bm\kappa \cdot \bm D  \cdot \bm \kappa }{2\rho}+\frac{R}{2} \frac{q ^2}{\rho^2} \right] \mathrm dV
 \label{eq:pcorr1}
\end{split}
\end{equation}
where $G=\mu/2\pi(1-\nu)$ (introduced for dimensional reasons),  $A$ and $R$ are dimensionless constants, $\bm D$ is a $2 \times 2$ dimensionless constant matrix \cite{zaiser2015local}, and $\rho_0=1/c^2b^2\gg \rho$ is a constant parameter with dislocation density dimension with $c$ being a constant determined by the core properties of the dislocation \cite{zaiser2015local}.  Similar expressions have been proposed by Hochrainer \cite{hochrainer2016thermodynamically} and Zaiser\cite{zaiser2015local}. We recall, however, that the plastic potential suggested here is not the coarse grained elastic energy derived by Zaiser\cite{zaiser2015local}, as is demonstrated for the 2D case in Appendix \ref{sec: plastic potential}. The second term in $P^\mathrm{corr}$ has to be invariant to the rotation of the Burgers vector, so
\begin{equation}
\begin{split}
 &\bm{\kappa}\cdot \bm{D} \cdot \bm{\kappa} = \\
 &\quad D_\mathrm{ss} (\bm{n}_\mathrm s \cdot \bm{\kappa})^2
 +D_\mathrm{ee} (\bm{n}_\mathrm e \cdot \bm{\kappa})^2+2D_\mathrm{se} (\bm{n}_\mathrm s \cdot \bm{\kappa})(\bm{n}_\mathrm e \cdot \bm{\kappa})
\end{split}
\end{equation}
where $D_\mathrm{ss}$, $D_\mathrm{ee}$, and $D_\mathrm{se}$ characterise correlations between screw dislocations, edge dislocations and between screw and enge dislocations, respectively, and form a symmetric positive definit tensor. Moreover, $\bm{n}_\mathrm s$, and  $\bm{n}_\mathrm e$ are unit vectors parallel and perpendicular (in the $xy$ plane) to the Burgers vector, respectively. 

What is really new as compared to the 2D case is the third term in Eq.~(\ref{eq:pcorr1}). Since the energy should not depend on the direction of the curvature (that is, the sign of $q$), as a lowest order approximation we suggest a quadratic form in $q$. Since the dimension of $q$ is m$^{-3}$, the parameter $R$ is dimensionless. With this, apart from $\rho_0$,  no material parameter with length dimension is introduced into the theory (the multiplier $G b^2$ cancels from the evolution equations). It should be mentioned, however, that the energy term related to curvature should account for line tension effects. Since line tension is a core effect it should go to infinity with decreasing dislocation core size. As a consequence one may suggest that $R$ is inversely proportional to $b^2 \rho$ (the core size is in the order of the size of the Burgers vector $b$). The issue requires further investigations.

According to Eqs.~(\ref{eq:taustar}) and (\ref{eq:P_split}) $\tau^*$ can be split into two parts, namely \begin{equation}
    \tau^* = \tau^\mathrm{mf} + \tau^\mathrm b
\end{equation}
with $\tau^\mathrm b$ being the back-stress. The latter is the contribution of the correlation part of the plastic potential $P^\mathrm{corr}$ as
\begin{equation}
\begin{split}
 \tau^\mathrm b&= \frac 1b \left[ \partial_y \frac{\delta P^\mathrm{corr}}{\delta \kappa_1}- \partial_x \frac{\delta P^\mathrm{corr}}{\delta \kappa_2} \right] \\
 &=G b\left[
 \partial_y\left(D_{11}\frac{\kappa_1}{\rho}+D_{12}\frac{\kappa_2}{2\rho}  \right)
\right. \\
 & \quad \left. -\partial_x\left(D_{22}\frac{\kappa_2}{\rho}+D_{12}\frac{\kappa_1}{2\rho}  \right)\right].  \label{eq:pcorr}
\end{split}
\end{equation}
If the system is not far from homogeneous, i.e.,\ $\kappa_{1,2}\ll \rho$ and $\partial_{x,y}\rho\ll \rho^{3/2}$ then
\begin{equation}
\begin{split}
 \tau^\mathrm b&=\frac{G b}{\rho}\left[\partial_y\left(D_{11}\kappa_1+D_{12}\kappa_2\right)
 \right. \\ & \quad \left .
 -\partial_x\left(D_{22}\kappa_2+D_{12}\kappa_1\right)\right],
\end{split}
\end{equation}
that corresponds to the straightforward generalization of the back-stress introduced for 2D\cite{groma2003spatial}, hence the identical notation.

Concerning the diffusion stresses $\tau^\mathrm d_1$ and $\tau^\mathrm d_2$ one obtains:
\begin{equation}
\begin{split}
\tau^\mathrm d_1 &= \frac 1{b \rho} \left[ \rho \, \partial_y \frac{\delta P^\mathrm{corr}}{\delta \rho} +q\,\partial_y \frac{\delta P^\mathrm{corr}}{\delta q} +\partial_y\left(\lambda_1 \rho \frac{\delta P^\mathrm{corr}}{\delta \rho} \right) \right] \\
&= \frac{Gb}{\rho} \left[ A \left(1+2\lambda_1+\lambda_1 \ln \frac{\rho}{\rho_0} \right) \partial_y \rho + R q \, \partial_y \frac{q}{\rho^2} \right],
\end{split}
\end{equation}
and similarly
\begin{equation}
\begin{split}
\tau^\mathrm d_2 &= -\frac 1{b \rho} \left[ \rho \, \partial_x \frac{\delta P^\mathrm{corr}}{\delta \rho} +q\,\partial_x \frac{\delta P^\mathrm{corr}}{\delta q} +\partial_x\left(\lambda_2 \rho \frac{\delta P^\mathrm{corr}}{\delta \rho} \right) \right] \\
&= -\frac{Gb}{\rho} \left[ A \left(1+2\lambda_2+\lambda_2 \ln \frac{\rho}{\rho_0} \right) \partial_x \rho + R q \, \partial_x \frac{q}{\rho^2} \right].
\end{split}
\end{equation}
Note that the constant $\rho_0$ does appear in the formulae, but only in terms also containing $\lambda_{1,2}$ as a multiplicative factor. We address this issue in the Discussion section below.

\section{Discussion}

The continuum theory of curved dislocations presented in the paper is a direct generalization of the 2D continuum theory of  straight parallel edge dislocations developed earlier by a systematic coarse-graining of the evolution equation of the individual dislocations \cite{groma2003spatial,groma2016dislocation}. But while the 2D continuum theory is directly linked to the discrete dislocation dynamics, building the direct link between the discrete and continuum description for the curved dislocation problem seems virtually impossible. Therefore, in order to have a closed theory  for the dependence of the velocities $v^\mathrm m$ and $v^\mathrm d_{1,2}$ on the dislocation state, one has to resort to phenomenological rules. In the current contribution we deduce these rules by closely following the 2D case, where the phenomenology was derived earlier to match the coarse grained theory. 

The proposed model  may be summarized as follows (see also Fig.~\ref{fig:sum}):
\begin{itemize}
 \item The state of the dislocation system is given by the fields: total dislocation density $\rho$, GND density vector $(\kappa_1,\kappa_2)$, and curvature density $q $.
 \item For the time evolution of these fields a ``dipole'' approximation is used leading to the Eqs.~(\ref{eq:trho},\ref{eq:tkappa1},\ref{eq:tkappa2},\ref{eq:tk}). 
 \item The dynamics of the system is obtained from a scalar functional $P[\bm{\chi},\rho,\kappa_1,\kappa_2,q ]$ called ``plastic potential''. In analogy to irreversible thermodynamics, the relevant quantities are the  appropriate combinations of the spatial derivatives of the different ``chemical'' potentials, which are the corresponding functional derivatives of the plastic potential. The key quantities are the ``effective stress'' $\tau^*$ that is the sum of the mean field and ``back'' stresses (Eqs.~(\ref{eq:taustar})), and the generalized ``diffusion'' stresses $\tau^\mathrm d_1$ and $\tau^\mathrm d_2$ which depend on the gradient of the dislocation density and the curvature field (Eqs.~(\ref{eq:tau1d},\ref{eq:tau2d})). 
\begin{figure}[!ht]
\begin{tcolorbox}[colframe=black!99!white,colback=white]
\color{red} Fields
\color{black} 
\begin{equation}
\rho(\bm{r},t),\ \kappa_1(\bm{r},t),\ \kappa_2(\bm{r},t),\ q(\bm{r},t) \nonumber
\end{equation}
\\ 
\color{red} Kinematics \color{black}
\begin{eqnarray}
 \partial_t \rho&=&-\partial_x(\rho v^\mathrm d_2)+\partial_y(\rho v^\mathrm d_1)+\partial_y(\kappa_1 v^\mathrm m)-\partial_x(\kappa_2 v^\mathrm m)
 \nonumber \\ &&+q v^\mathrm m+ \lambda_1\rho\partial_y v^\mathrm d_1- \lambda_2 \rho\partial_x v^\mathrm d_2 \nonumber \\
 \partial_t \kappa_1&=&\partial_y(\rho v^\mathrm m+\kappa_1 v^\mathrm d_1+\kappa_2 v^\mathrm d_2) \nonumber \\
 \partial_t \kappa_2&=&-\partial_x(\rho v^\mathrm m+\kappa_1 v^\mathrm d_1+\kappa_2 v^\mathrm d_2) \nonumber \\
 \partial_t q &=&-\partial_x\left(q v^\mathrm d_2-v^\mathrm m Q_1\right) 
 +\partial_y\left(q v^\mathrm d_1+v^\mathrm m Q_2\right) \nonumber
\end{eqnarray}
\color{red} Dynamics \color{black} \\ \vspace{8pt}
Plastic potential: $P[\bm{\chi},\rho,\kappa_1,\kappa_2,q ]$ \\ \color{red} $\Downarrow$ \color{black} \\
$\tau^*$, $\tau^\mathrm d_1$, $\tau^\mathrm d_2$ \\ \vspace{10pt}
\color{red} Velocities \color{black} \\ \vspace{10pt}
\includegraphics[angle=0,width=7.5cm]{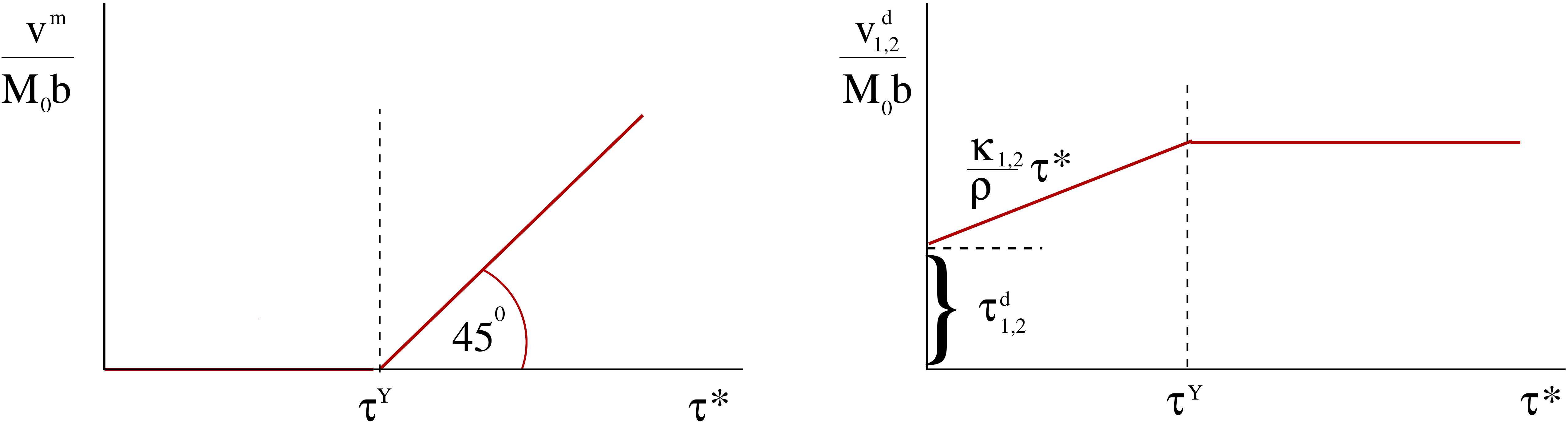}
\end{tcolorbox}
\caption{Summary of the model} \label{fig:sum}
\end{figure}
 \item The $\tau^*$ and $\tau^\mathrm d_{1,2}$ dependence of the velocity fields $v^\mathrm m$ and $v^\mathrm d_{1,2}$ are indicated in Fig.~\ref{fig:vel}. Below the flow stress the mean velocity  $v^\mathrm m$ vanishes, while above it increases linearly with $\tau^*$. In the non-flowing regime the velocities  $v^\mathrm d_{1,2}$ are linear in $\tau^*$, whereas in the flowing one they remain constant upon increasing stress.
\end{itemize}

A central but yet qualitative finding of the current study is that the requirement of thermodynamic consistency may yield a state-dependent Taylor factor $\alpha$, at least for inhomogeneous and strongly curved dislocation configurations. Such corrections to the `Taylor law' are well-known in the literature on work-hardening \cite{Basinski_b79}, but they have not been derived from a continuum dislocation theory before. 

Going from systems of straight parallel edge dislocations towards the current single slip situation with curved dislocations is a small but important step towards a more general dislocation-based theory of plasticity. Before discussing a few aspects of generalizing the theory, however, we need to address an aspect, which might seem to refute the current results: 
it is found by numerous experimental investigations that most macroscopic properties of dislocation systems, like flow stress, and the length scale of dislocation patterns, scale with the average dislocation-dislocation spacing $l_\mathrm c=1/\sqrt{\rho}$. This means, systems with different dislocation densities can be scaled to each other with the  scaling factor $l_\mathrm c$. This is called the ``principle of similitude'' \cite{zaiser2006scale,zaiser_s14}.

In order to fulfill this principle, a continuum theory of dislocations may not contain parameters (or combinations of parameters) with length dimension other than $l_\mathrm c$. The plastic potential given by Eq.~(\ref{eq:pcorr}), however, contains the parameter $\rho_0$ with 1/length$^2$ dimension. Consequently, at first sight the proposed plastic potential, $P_\mathrm{corr}$ violates the principle of similitude. Concerning the evolution equations, however, $\rho_0$ appears only in the first term of $B^\mathrm m$ and in the last term for $\tau^\mathrm d_1$ and $\tau^\mathrm d_2$, cf.\ Eqs.~(\ref{eq:b0},\ref{eq:tau1d}, and \ref{eq:tau2d}), respectively. As it is discussed above $B^\mathrm m$ can be absorbed into the $\alpha$ parameter of the flow stress. For nearly homogeneous systems with small curvature $\alpha$ is independent from $B^\mathrm m$. Concerning $\tau^\mathrm d_1$ and $\tau^\mathrm d_2$ they have a logarithmic $\rho_0$ dependence in the term proportional to $\lambda_{1,2}$. Such logarithmic deviations from the principle of similitude are well-known to occur due to curvature effects \cite{zaiser_s14}.

Regarding the derivation of a more general theory of plasticity we note two simplifying assumptions which will need to be relaxed in the future:

I. In the above considerations the mobility $M_0$  is modelled as constant, resulting in a linear relation between the velocity $v^\mathrm m$ and the stress. This may be generalized by allowing that $M_0$ depends on the stress. Taking the form
\begin{eqnarray}
 M_0(\tau^*)=M^* f\left(\frac{\tau^*-\tau^\mathrm y}{\tau^\mathrm y} \right),
\end{eqnarray}
where $M^*$ is a positive constant and $f(x)$ is any non-negative function, does in general comply with inequality (\ref{eq:pp}). Common phenomenological plasticity theories with a power function relation between plastic shear rate and shear stress can be recovered if $f(x)$ is chosen as a power function.

II. Formally it appears relatively straightforward to generalize the theory for multiple slip. In this case, however, local dislocation-dislocation interactions, like formation of junctions and dislocation annihilation, cannot be neglected. Moreover, cross slip and at higher temperature climb may play an important role, too. These phenomena need to be incorporated into the proposed theory as source and sink terms. Promising, though mostly phenomenological ways, how to incorporate some of these phenomena in continuum dislocation theories were recently proposed by El-Azab \emph{et al.}\cite{el2006statistical,xia2015computational,lin2020implementation} and Schulz \emph{et al.}\cite{schulz2017mechanism,sudmanns2019dislocation}. These extensions proved successful in describing various technologically important situations, such as torsion of microwires and compression of micropillars \cite{ZOLLER2020198, zoller2020microstructure}.

As a concluding remark we can state that the proposed continuum theory of curved dislocations is established in a systematic manner based on the continuum theory of straight parallel dislocations. While this is an important step toward a general theory of crystal plasticity, a wealth of issues still need to be addressed in future research. 

\appendix

\section{Derivation of Eq.~(\ref{eq:cosrho})}
The detailed derivation of the identity (\ref{eq:cosrho})
\begin{equation}
\begin{split}
&\cos \varphi \left(\hat{\text{Div}}(\rho' {\bm V})+q'v'\right) \\
& \quad =\partial_y(\rho' v')-\partial_\varphi\left(-q'v'\sin \varphi-\rho'\hat{L}(v') \cos \varphi\right) 
\end{split}
\end{equation}
is given below. We use two conditions: (i) from equation $\hat{\text{Div}}(\rho' \bm L)=0$ one gets that
\begin{eqnarray}
 \partial_\varphi q'=-\cos \varphi \, \partial_x \rho'-\sin \varphi \, \partial_y \rho'=: - \hat{l}(\rho'),
\end{eqnarray}
where the operator $\hat l = \cos\varphi \, \partial_x + \sin \varphi \, \partial_y$ was introduced, and (ii) from Eq.~(\ref{eq:L}) one obtains that
\begin{equation}
\begin{split}
 \vartheta' &:= -\hat{L}(v')=-\hat{l}(v')-k'\partial_\varphi v' \\
 &= -\cos \varphi \, \partial_x v' - \sin \varphi \, \partial_y v'-k'\partial_\varphi v'.
\end{split}
\end{equation}
With the above equations one can find that
\begin{equation}
\begin{split}
 & \cos \varphi \left(\hat{\text{Div}}(\rho' \bm V) + q'v'\right) \\
 &\quad = [\cos \varphi \, \partial_y(\rho' v')-\sin \varphi \, \partial_x(\rho' v')-\partial_\varphi(\rho' \vartheta')]\cos\varphi \\
 &\qquad+q' v'\cos\varphi \\
 &\quad=\cos^2 \varphi \, \partial_y(\rho' v')-\sin \varphi  \cos \varphi \, \partial_x(\rho' v') \\
&\qquad- \partial_\varphi(\rho' \vartheta')\cos\varphi + q' v' \cos\varphi \\
&\quad=\partial_y(\rho' v')-[\cos \varphi \, \partial_x(\rho' v')+\sin \varphi \, \partial_y(\rho' v')]\sin \varphi \\
&\qquad +q' v' \cos\varphi -\partial_\varphi(\rho' \vartheta')\cos\varphi \\
&\quad=\partial_y(\rho' v')-\hat{l}(\rho' v')+q' v'\cos\varphi - \partial_\varphi(\rho' \vartheta')\cos\varphi \\
&\quad=\partial_y(\rho' v')-\hat{l}(\rho' v')+\rho'\hat{L}(v')\sin\varphi +q' v'\cos\varphi \\
&\qquad-\partial_\varphi(\rho' \vartheta')\cos\varphi+\rho' \vartheta' \sin \varphi \\
&\quad=\partial_y(\rho' v')-[\hat{l}(\rho' v')-\rho' \hat{l}(v')]\sin \varphi
+q' \partial_\varphi v' \sin \varphi \\
&\qquad +q' v' \cos \varphi-\partial_\varphi(\rho' \vartheta')\cos \varphi +\rho' \vartheta'\sin \varphi \\
&\quad=\partial_y(\rho' v')-\hat{l}(\rho')v' \sin \varphi
+q' \partial_\varphi v' \sin \varphi +q'v' \cos \varphi \\
&\qquad -\partial_\varphi(\rho' \vartheta')\cos \varphi +\rho' \vartheta' \sin \varphi \\
&\quad= \partial_y(\rho' v')+\partial_\varphi(q') v'\sin \varphi
+q' \partial_\varphi v' \sin \varphi +q'v' \cos \varphi \\
&\qquad-\partial_\varphi(\rho' \vartheta')\cos \varphi +\rho' \vartheta'\sin \varphi \\
&\quad= \partial_y(\rho' v')-\partial_\varphi\left(-q'v'\sin \varphi+\vartheta'\rho' \cos \varphi\right)
\end{split}
\end{equation}

\section{Derivation of Eq.~(\ref{eq:q})}

For the derivation of Eq.~(\ref{eq:q}) we note that Eq.~\eqref{eq:k} was originally derived from the evolution of $q'$ in the form \cite{hochrainer2007three}
\begin{eqnarray}
\partial_t q'&=&-k \hat{\text{Div}}(\rho' \bm V) -  \rho' \hat{L}(\hat{L} (v'))-\rho'\hat{V}(k'). \label{eq:q0}
\end{eqnarray}
From this form Eq.~(\ref{eq:q}) follows by
\begin{equation}
\begin{split}
 \partial_t q'&=-k \hat{\text{Div}}(\rho' \bm V) -  \rho' \hat{L}(\hat{L} (v'))-\rho'\hat{V}(k') \\
 &=-\hat{\text{Div}}(\rho' k' \bm V)+ \rho' \hat{V}(k') -\rho' \hat{V}(k')+\rho' \hat{L}(\vartheta') \\
 &=-\hat{\text{Div}}(\rho' k' \bm V)+ \hat{\text{Div}}(\rho' \vartheta' \bm L) - \vartheta'
 \hat{\text{Div}}(\rho'\bm L) \\
  &=-\hat{\text{Div}}(\rho' k' \bm V-\rho'\vartheta'\bm L ) \\
  &=-\hat{\text{Div}}(q' \bm V + \rho'\hat{L} (v') \bm L ),
\end{split}
\end{equation}
where we used the product rule
\begin{eqnarray}
 \hat{\text{Div}}(f' \bm X)=\hat{X}(f') + f'  \hat{\text{Div}}( \bm X)
\end{eqnarray}
in which $f'$ is a scalar function, $\bm{X}=(X_x,X_y,X_\varphi)$ is a  vector function, and $\hat{X}$ denotes the directional derivative operator $\hat{X}=X_x\partial_x+X_y\partial_y+X_\varphi\partial_\varphi$.

\section{Mean-field stress}
\label{sec:mean_field}
In this appendix we consider the quantity
\begin{eqnarray}
\frac1b \left( \partial_y \frac{\delta P^\mathrm{mf}}{\delta \kappa_1}- \partial_x \frac{\delta P^\mathrm{mf}}{\delta \kappa_2} \right).
\end{eqnarray}
The only term in $ P^\mathrm{mf}$ that depends on the GND variables $\kappa_1$ and $\kappa_2$ will be denoted as $P^\mathrm p$:
\begin{eqnarray}
 P^\mathrm p=\int \chi_{ij} \eta_{ji} \mathrm dV.
\end{eqnarray}
Since the incompatibility tensor is $\eta_{ij}=-(\text{Inc}\,  \bm \epsilon^\mathrm p)_{ij}$ where $\bm \epsilon^\mathrm p$ is the plastic deformation tensor, with partial integration one obtains that
\begin{eqnarray}
 P^\mathrm p=-\int \sigma_{ij} \epsilon^\mathrm p_{ji} \, \mathrm dV
\end{eqnarray}
where we used that $\bm \sigma=\mathrm{Inc}\, \bm\chi$. Without restricting the generality we can assume that  the Burgers vector is parallel to the $x$ axis ($\bm b=(b,0,0)$). Since the slip normal is parallel to the $z$ axis, in this case only the $31$ component of $\bm \beta^\mathrm p$ differs from zero. With this 
\begin{eqnarray}
 P^\mathrm p=-\int \sigma_{13} \beta^\mathrm p_{31} \, \mathrm dV.
\end{eqnarray}
One can now introduce $S_1$ and $S_2$ scalar fields such that $\sigma_{13}=-\partial_y S_1+\partial_x S_2$ where $S_{1,2}$ are not uniquely defined but as it will be shown below this does not influence the final result. With Eqs.~(\ref{eq:k1},\ref{eq:k2}) again after partial integration
\begin{eqnarray}
 P^\mathrm p=-\int (S_1 \kappa_1+S_2\kappa_2)b \, \mathrm dV.
\end{eqnarray}
From this one arrives at 
\begin{eqnarray}
 \frac 1b \left( \partial_y \frac{\delta P^\mathrm{mf}}{\delta \kappa_1}- \partial_x \frac{\delta P^\mathrm{mf}}{\delta \kappa_2} \right)=\sigma_{13} = \tau^\mathrm{mf}.
\end{eqnarray}

\section{Plastic potential}\label{sec: plastic potential}
Since the issue has not been published earlier we give a short discussion below why one has to distinguish between the coarse grained elastic energy and the ``plastic potential'' proposed to used for giving the velocities $v^\mathrm m$ and $v^{\mathrm d}_{1,2}$. We discuss only the 2D straight edge dislocation evolution problem. The 3D generalization is far from straightforward (see Ref.~\onlinecite{zaiser2015local}). 

Let us start with the equation of motion of the individual dislocations. We consider a system of parallel edge dislocations with Burgers vector $\bm {b}=(\pm b,0,0)$ and line direction $\bm{l}=(0,l,0)$. Assuming overdamped motion, the velocity of the {\it i}th dislocation is proportional to the force acting on it: 
\begin{eqnarray}
\frac{\mathrm d x_i}{\mathrm d t}=M_0 b \left(\sum_{j=1}^N s_i s_j \tau_\mathrm{ind}(\bm{r}_i-\bm{r}_j) \right) \label{eq:eqm}						
\end{eqnarray}
where $\bm{r}_i$ is the position of the {\it i}th dislocation in the $xz$ plane, $s_i=\pm 1$ is the sign of the {\it i}th dislocation, $N$ is the number of dislocations, $M_0$ is a mobility constant, and  $\tau_\mathrm{ind}(\vec{r})$ is the shear stress generated by a dislocation with positive sign. For simplicity, no external load is considered here, but it can be added in a straightforward manner.   

Since the shear stress is the appropriate second derivative of the stress potential \cite{kroner1981continuum}
\begin{eqnarray}
 \tau_\mathrm{ind}=\partial_x\partial_z \chi_\mathrm{ind}
\end{eqnarray}
it can be given as $\tau_\mathrm{ind}=-\partial_x \Phi_\mathrm{ind}$ where $\Phi_\mathrm{ind}=-\partial_z \chi_\mathrm{ind}$. This means that $b\Phi_\mathrm{ind}$ is the elastic interaction energy between two dislocations with the same sign. 
It follows that the equation of the motion of the {\it i}th dislocation (\ref{eq:eqm}) can be given as
\begin{equation}
\frac{\mathrm d x_i}{\mathrm d t}=-M_0\partial_{x_i} V(\{\bm{r}_i\}) \label{eq:eqm2}
\end{equation}
where
\begin{equation}
 V(\{\bm{r}_i\})=b\sum_{i,j} s_is_j \Phi_(\mathrm{ind}( \bm{r}_i-\bm{r}_j)
\end{equation}
is the total elastic interaction energy per unit length.  

It is easy to see that due to the dissipative dislocation motion
\begin{eqnarray}
 \frac{\mathrm d}{\mathrm dt} V(\{\bm{r}_i\})=-\frac{1}{M_0} \sum_{i=1}^N \left(\frac{\mathrm dx_i}{\mathrm dt}\right)^2\leqq 0
\end{eqnarray}
so the total ``discrete'' elastic energy cannot increase during the evolution of the dislocation system.

In order to have a continuum theory derived from the evolution of the individual dislocations one can perform coarse-graining at two different ways indicated in Fig.~\ref{fig:cg}.  
\begin{widetext}\begin{center}
\begin{figure}[!ht]
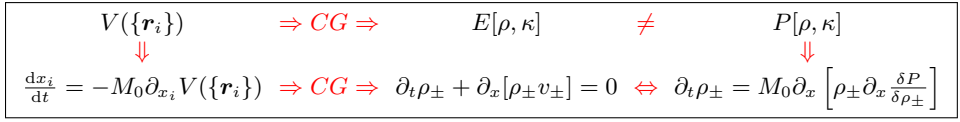
 
\begin{equation}\boxed{
\begin{array}{ccccc}
 V(\{\bm{r}_i\}) & \color{red} \Rightarrow CG \Rightarrow & E[\rho, \kappa] & \color{red} \neq &  P[\rho, \kappa] \\
 \color{red}\Downarrow & & & & \color{red} \Downarrow \\
 \frac{\mathrm dx_i}{\mathrm dt}=-M_0\partial_{x_i} V(\{\bm{r}_i\}) & \color{red}\Rightarrow CG \Rightarrow & \partial_t\rho_{\pm}+\partial_x[\rho_{\pm} v_{\pm}]=0& \color{red}\Leftrightarrow& \partial_t\rho_{\pm}=M_0\partial_x\left[\rho_{\pm} \partial_x\frac{\delta P}{\delta \rho_\pm}\right] \nonumber
\end{array}}
\end{equation}
\caption{Coarse-graining (CG) at different levels.} \label{fig:cg}
\end{figure}\end{center}
\end{widetext}
As it is explained in detail by Zaiser in Ref.~\onlinecite{zaiser2015local} a coarse-grained energy $E[\kappa,\rho]$ can be obtained from $V(\{\bm{r}_i\})$ with the knowledge of the dislocation-dislocation correlation functions. 

An alternative way is a systematic coarse-graining of the system of the equation of motion of dislocations (\ref{eq:eqm2}). For the detailed derivation see Ref.~\onlinecite{groma2003spatial,groma2016dislocation}. As it is explained in Ref.~\onlinecite{groma2016dislocation} (see also above) the evolution equations for the fields $\rho_{\pm}$ can be obtained from a scalar functional $P[\rho, \kappa]$ called ``plastic potential'' that is directly derived from the equation of motion of dislocations. If we, however, compare $E[\rho, \kappa]$ and $P[\rho, \kappa]$ we find that the functional form of the two quantities are the same, namely
\begin{eqnarray}
 P_\mathrm{mf}+\int G b^2 \left[A \rho \ln\left(\frac{\rho}{\rho_0}\right)+ 
 \frac{\kappa D \kappa}{2\rho}\right]\mathrm dx \mathrm dy \label{eq:pcorr2D}
\end{eqnarray}
but the parameters $A$ and $D$ appearing in them, determined by the dislocation-dislocation correlation functions, are different \cite{zaiser2015local,groma2016dislocation}.
In the local density approximation \cite{zaiser2015local,groma2016dislocation} for the plastic potential
\begin{equation}
\begin{split}
 D_\mathrm{pp}&=\frac{\rho}{Gb}\int \left[(d^\mathrm s+d^\mathrm d)x \partial_x \Phi_\mathrm{ind}  \right]\mathrm dx \mathrm dy \\
 A_\mathrm{pp}&=\frac{\rho}{Gb}\int \left[(d^\mathrm s-d^\mathrm d)x \partial_x \Phi_\mathrm{ind}  \right]\mathrm dx\mathrm dy
\end{split}
\end{equation}
while for the coarse grained energy
\begin{equation}
\begin{split}
 D_\mathrm{cge}&=\frac{\rho}{Gb}\int \left[(d^\mathrm s+d^\mathrm d)\Phi_\mathrm{ind}  \right]\mathrm dx \mathrm dy \\
 A_\mathrm{cge}&=-\frac{1}{4}
\end{split}
\end{equation}
where $d^\mathrm s$ and $d^\mathrm d$ are spatial correlation functions between dislocations of the same and the opposite signs, respectively. It should be noted that for the coarse grained energy $A$ is independent from the correlation function that is obviously cannot lead to the right evolution equation where all the terms but the mean-field stress are related to dislocation-dislocation correlations.  

Since $P[\rho, \kappa]$ is obtained from the equation of motion of the individual dislocations and, as it is explained above, it cannot increase during the evolution of the coarse-grained fields, the plastic potential is the quantity we have to use in the generalized 3D theory, too.

\begin{acknowledgments}
  This work has been supported by the National Research, Development and Innovation Office of Hungary (PDI and IG, project Nos.~NKFIH-K-119561) and the ELTE Institutional Excellence Program (TKP2020-IKA-05) supported by the Hungarian Ministry of Human Capacities.
  \end{acknowledgments}

%

\end{document}